\documentclass[aps,prd,amsmath,amssymb,preprintnumbers,nofootinbib,a4paper,preprint]{revtex4-1}
\pdfoutput=1
\usepackage{amsthm}
\usepackage{graphicx}
\usepackage{color}
\newcommand{\beq}{\begin{eqnarray}}
\newcommand{\eeq}{\end{eqnarray}}

\newcommand{\bmp}{\noindent\begin{minipage}{16cm}}
\newcommand{\emp}{\end{minipage}\vskip 7mm} 

\usepackage{dcolumn}
\usepackage{bm}
\usepackage{bbm}
\usepackage{subfigure}
\usepackage{pxfonts}

\theoremstyle{definition}

\theoremstyle{plain}

\usepackage{epsfig}

\usepackage{graphicx}
\usepackage{multirow}
\usepackage[margin=5pt, font=normalsize,labelfont=bf,format=plain,justification=centerlast]{caption}

\definecolor{rossoCP3}{cmyk}{0,.88,.77,.40}

\def\lsim{\mathrel{\rlap{\lower4pt\hbox{\hskip1pt$\sim$}}
    \raise1pt\hbox{$<$}}}                
\def\gsim{\mathrel{\rlap{\lower4pt\hbox{\hskip1pt$\sim$}}
    \raise1pt\hbox{$>$}}}                

\newcommand{\be}{\begin{eqnarray}}
\newcommand{\ee}{\end{eqnarray}}

\usepackage[
bookmarks]{hyperref}
\usepackage{color}
\hypersetup{colorlinks, bookmarksopen, bookmarksnumbered,
citecolor=verdes, linkcolor=bluc, pdfstartview=FitH, urlcolor=rossos}

\definecolor{grigio}{cmyk}{0,0,0,0.1}
\definecolor{rosa}{cmyk}{0,0.1,0.1,0.02}
\definecolor{rosino}{cmyk}{0,0.05,0.05,0.02}
\definecolor{rosas}{cmyk}{0,0.3,0.25,0.05}
\definecolor{celeste}{cmyk}{0.1,0,0,0.02}
\definecolor{giallino}{cmyk}{0,0,0.1,0.02}
\definecolor{rosso}{cmyk}{0,1,1,0.4}
\definecolor{rossos}{cmyk}{0,1,1,0.55}
\definecolor{rossoc}{cmyk}{0,1,1,0.2}
\definecolor{blu}{cmyk}{1,1,0,0.3}
\definecolor{blus}{cmyk}{1,1,0,0.5}
\definecolor{bluc}{cmyk}{1,1,0,0.1}
\definecolor{blucc}{cmyk}{0.7,0.5,0,0}
\definecolor{viola}{cmyk}{0,1,0,0.6}
\definecolor{viola2}{cmyk}{0,1,0.2,0.6}
\definecolor{verde}{cmyk}{0.92,0,0.59,0.25}
\definecolor{verdec}{cmyk}{0.92,0,0.59,0.15}
\definecolor{verdes}{cmyk}{0.92,0,0.59,0.4}
\definecolor{verdino}{cmyk}{0.12,0,0.09,0.02}
\definecolor{giallo}{cmyk}{0,0,1,0}
\definecolor{gialloverde}{cmyk}{0.44,0,0.74,0}
\definecolor{Titolo}{rgb}{0.752941176,0.576470588,0.992156863}
\definecolor{altro}{rgb}{0.094117647,0.650980392,0.643137255}
\definecolor{Peanuts}{rgb}{0.2, 0.4, 0.6}
\definecolor{Pean1}{rgb}{0.6, 0.8, 0.4}
\definecolor{BHO}{rgb}{0.2, 0.8, 1}
\definecolor{Daria}{rgb}{0, 0.9412, 0}
\definecolor{UniPi}{rgb}{0.2549, 0.4627, 0.6275}
\definecolor{UniPidue}{rgb}{0.3216, 0.5804, 0.7882}

\newcommand{\ud}{\text{d}}
\usepackage{youngtab}
\usepackage{slashed}

\definecolor{rossoCP3}{cmyk}{0,.88,.77,.40}

\begin{document}
\title{\Large  \color{rossoCP3}   
 Composite Goldstone Dark Matter: Experimental Predictions from the Lattice}
\author{Ari Hietanen$^{\color{rossoCP3}{\varheartsuit}}$}\email{hietanen@cp3-origins.net} 
\author{Randy Lewis$^{\color{rossoCP3}{\spadesuit}}$}\email{randy.lewis@yorku.ca}
\author{Claudio Pica$^{\color{rossoCP3}{\varheartsuit}}$}\email{pica@cp3-origins.net} 
\author{Francesco Sannino$^{\color{rossoCP3}{\varheartsuit}}$}\email{sannino@cp3-origins.net} 

\affiliation{
\vspace{5mm}
{$^{\color{rossoCP3}{\varheartsuit}}${ \color{rossoCP3}  \rm CP}$^{\color{rossoCP3} \bf 3}${\color{rossoCP3}\rm-Origins} \& the {\color{rossoCP3} \rm Danish IAS},
University of Southern Denmark, Campusvej 55, DK-5230 Odense M, Denmark}
\vspace{5mm} \\
\mbox{ $^{\color{rossoCP3}{\spadesuit}}$Department of Physics and Astronomy,  {\color{rossoCP3} York University}, Toronto, M3J 1P3, Canada}
\vspace{1cm}
}

\begin{abstract}
We study, via first principles lattice simulations, the nonperturbative dynamics of $SU(2)$ gauge theory with two fundamental Dirac flavors. The model can be used simultaneously as a template for composite Goldstone boson  dark matter and for breaking the electroweak symmetry dynamically. We compute the form factor, allowing us to estimate the associated electromagnetic charge radius. Interestingly we observe that the form factor obeys vector meson dominance even for the two color theory. We finally compare the model predictions with dark matter direct detection experiments. We find that the composite Goldstone boson dark matter cross sections is constrained by the most stringent direct-detection experiments. Our results are a foundation for quantitative new composite dynamics relevant for model building, and are of interest to current experiments.
 \\  
[.1cm]
{\footnotesize  \it Preprint: CP$^3$-Origins-2013-30 DNRF90 \& DIAS-2013-30}
\end{abstract}

\maketitle

\section{Introduction}
Unveiling the nature of dark matter (DM) constitutes a fundamental problem in physics.  DM plays an important role in large scale structure formation as well as the evolution of the Universe. Several earth and space based experiments are searching for DM to study its properties.   

Starting from the simple observation that the bulk of the ordinary matter is composite, i.e. is made by neutrons and protons, it is justified and intriguing to explore the paradigm according to which also DM has a composite nature.  

Composite Higgs models, such as the contemporary Technicolor models,
present relevant examples in which the model can simultaneously
address the naturalness problems of the SM and offer well-motivated
composite DM states. Composite DM states in these models can be heavy,
typically of the order a few TeVs
\cite{Nussinov:1985xr,Dietrich:2006cm,Nardi:2008ix}, when identified
with the composite fermions of the theory, or light, i.e.\ with masses
ranging from a few GeVs to hundreds of GeVs if identified with the
(pseudo) Goldstone \cite{Gudnason:2006ug}. Several
asymmetric DM candidates appeared in the literature
~\cite{Nussinov:1985xr,Gudnason:2006ug,
  Foadi:2008qv,Khlopov:2008ty,Dietrich:2006cm,Sannino:2009za,Ryttov:2008xe,Nardi:2008ix,Kaplan:2009ag,Frandsen:2009mi,Belyaev:2010kp}. An
interesting variation on the main composite Higgs theme is the one
according to which the composite Higgs is also a (pseudo) Goldstone
boson by Kaplan and Georgi \cite{Kaplan:1983fs}. A unified description
of composite Higgs models is given in \cite{Cacciapaglia:2014uja}.

However, so far, composite Goldstone DM phenomenology relied solely on the symmetries of the underlying gauge theory and effective Lagrangians descriptions. While these approaches are useful, a first principle estimate of the form factors dictating the interactions, and associated physics, between the DM candidate and ordinary matter is essential to guide the experimental searches. Furthermore, due to the composite nature of the DM states, the knowledge of the energy dependence of the form factors allows to study and relate the DM properties in different energy regimes ranging from a few KeVs to hundreds of GeVs. 

Here we consider a template of composite Goldstone boson DM  \cite{Ryttov:2008xe} investigated on the lattice in \cite{Lewis:2011zb,Hietanen:2014xca}, namely an SU(2) gauge theory with two fundamental
fermion flavors.  We view this theory as the kernel from which more
elaborate models can grow.  For example, there are extensions that show how
a 125 GeV scalar can emerge \cite{Foadi:2012bb}.
The USQCD collaboration highlighted this lattice theory in a recent white paper \cite{Appelquist:2013sia} and studied the effects of additional fermions in \cite{Appelquist:2013pqa}.
Other groups have reported results at strong coupling \cite{Nagai:2009ip}, results with nonzero chemical potential \cite{Hands:2006ve,Hands:2007uc,Hands:2010gd,Hands:2011hd,Cotter:2012mb}, and results with chiral lattice fermions \cite{Matsufuru:2014dva}.  Dark matter candidates from nuclei in this lattice theory were discussed in \cite{Detmold:2014qqa,Detmold:2014kba}.  Dark matter in the related SU(3) and SU(4) lattice theories were considered in \cite{Appelquist:2013ms} and \cite{Appelquist:2014jch} respectively.
Our minimal template has the appeal to address simultaneously electroweak symmetry breaking and the origin of a naturally-light DM candidate  \cite{Ryttov:2008xe}.

The template is an SU(2) gauge theory with two fundamental fermion flavors,
named $u$ and $d$.
This action has a global SU(4) symmetry, and the lattice simulations of
Ref.~\cite{Lewis:2011zb} showed that it is dynamically broken to Sp(4),
thereby producing five Goldstone bosons.
Three of these are eaten by the $W^\pm$ and $Z$ bosons; the remaining pair of
Goldstones is the DM candidate and its antiparticle.
Depending on the cross section for annihilation into standard model fields one can have a symmetric  (i.e.\ thermal relic density),  asymmetric, or a mixed scenario \cite{Belyaev:2010kp}. 
An exact Goldstone boson would be massless but, like the pions of QCD, the
DM candidate can acquire a small mass from explicit symmetry
breaking through new interactions breaking the original $SU(4)$ symmetry to $SU_L(2) \times SU_R(2)\times U(1)$ while keeping the $u$ and $d$ massless. The  effective Lagrangian operator was constructed in \cite{Ryttov:2008xe} and corresponds to an effective four-fermion interaction. However, as recently pointed out in \cite{Cacciapaglia:2014uja}, standard model radiative corrections alone are sufficient to give mass to the would be Goldstone Boson. The present model allows us to study the interaction between composite DM and ordinary matter by determining the associated electric dipole moment.

The light DM limit was originally introduced to explore models of interfering DM \cite{Foadi:2008qv,DelNobile:2011je,DelNobile:2011yb} useful to alleviate the tension between the experimental observations by DAMA/LIBRA \cite{Bernabei:2008yi} and the limits set by XENON100 \cite{Aprile:2011hi,Aprile:2012nq} and CDMS \cite{Ahmed:2009zw}. However with the very constraining results by LUX \cite{Akerib:2013tjd} it has become increasingly harder to reconcile these anomalies. We will therefore assume here a very conservative attitude and compare our results only with the most severe exclusion results from LUX, XENON100 and SuperCDMS \cite{Agnese:2014aze}.

The present paper is organized as follows.
Section~\ref{sec:lattheory}
explains how lattice computations of the form factors can be
performed.
Section~\ref{sec:relations} derives relationships among the form factors of
the five Goldstone bosons.
Section~\ref{sec:latresults}
presents the numerical results of our lattice simulations
and demonstrates that vector meson saturation for the form factors applies even in the case of the two color theory.
Section~\ref{sec:phenoresults} combines the lattice results to determine the electroweak form factor and the associated DM proton cross section. The effect of Higgs exchange and the direct comparison with the experimental data is presented in Section~\ref{sec:experiments}.  Section~\ref{sec:conc} contains our conclusions.

\section{The Lattice Method}\label{sec:lattheory}

In the continuum, the Lagrangian for our technicolor template is
\begin{equation}
{\cal L} = -\frac{1}{4}F_{\mu\nu}^aF^{a\mu\nu}
         + \overline{u}(i\gamma^\mu D_\mu-m_u)u
         + \overline{d}(i\gamma^\mu D_\mu-m_d)d
\end{equation}
which can be discretized in the familiar way to arrive at a Wilson action,
\begin{eqnarray}
S_W &=& \frac{\beta}{2}\sum_{x,\mu,\nu}\left(1-\frac{1}{2}{\rm ReTr}U_\mu(x)
        U_\nu(x+\hat\mu)U_\mu^\dagger(x+\hat\nu)U_\nu^\dagger(x)\right)
      + \sum_x\overline{\psi}(x)(4+m_0)\psi(x) \nonumber \\
   && - \frac{1}{2}\sum_{x,\mu}\left(\overline{\psi}(x)(1-\gamma_\mu)U_\mu(x)\psi(x+\hat\mu)
   +\overline{\psi}(x+\hat\mu)(1+\gamma_\mu)U_\mu^\dagger(x)\psi(x)\right) \,, 
\end{eqnarray}
where $U_\mu$ is the gauge field and $\beta$ the gauge coupling in conventional
lattice notation.  $\psi$ is the doublet of $u$ and $d$ fermions, and $m_0$ is
the 2$\times$2 diagonal mass matrix.

Mesons will couple to local operators of the form
\begin{eqnarray}
{\cal O}_{\overline{u}d}^{(\Gamma)}(x) &=& \overline{u}(x)\Gamma d(x) \,, \\
{\cal O}_{\overline{d}u}^{(\Gamma)}(x) &=& \overline{d}(x)\Gamma u(x) \,, \\
{\cal O}_{\overline{u}u\pm\overline{d}d}^{(\Gamma)}(x)
    &=& \frac{1}{\sqrt{2}}\bigg(\overline{u}(x)\Gamma u(x)
        \pm \overline{d}(x)\Gamma d(x)\bigg) \,,
\end{eqnarray}
where $\Gamma$ denotes any product of Dirac matrices.
Baryons (which are diquarks in this two-color theory) will couple to local
operators of the form
\begin{eqnarray}
{\cal O}_{ud}^{(\Gamma)}(x) &=& u^T(x)(-i\sigma^2)C\Gamma d(x) \,, \\
{\cal O}_{du}^{(\Gamma)}(x) &=& d^T(x)(-i\sigma^2)C\Gamma u(x) \,, \\
{\cal O}_{uu\pm dd}^{(\Gamma)}(x)
    &=& \frac{1}{\sqrt{2}}\bigg(u^T(x)(-i\sigma^2)C\Gamma u(x)
        \pm d^T(x)(-i\sigma^2)C\Gamma d(x)\bigg) \,,
\end{eqnarray}
where the Pauli structure $-i\sigma^2$ acts on color indices while the
charge conjugation operator $C$ acts on Dirac indices.

A photon can couple to a local vector operator such as
${\cal O}_{\overline{u}u\pm\overline{d}d}^{(\gamma_\mu)}$ which becomes a
conserved current in the continuum limit but is not conserved in the lattice
theory.  In studies of the electroweak form factors, it is advantageous
to work directly with the lattice conserved currents,
\begin{eqnarray}
V_\mu^u(x)
&=& \frac{1}{2}\overline{u}(x+\hat\mu)(1+\gamma_\mu)U_\mu^\dagger(x)u(x)
  - \frac{1}{2}\overline{u}(x)(1-\gamma_\mu)U_\mu(x)u(x+\hat\mu) \,, \\
V_\mu^d(x)
&=& \frac{1}{2}\overline{d}(x+\hat\mu)(1+\gamma_\mu)U_\mu^\dagger(x)d(x)
  - \frac{1}{2}\overline{d}(x)(1-\gamma_\mu)U_\mu(x)d(x+\hat\mu) \,,
\end{eqnarray}
that are easily combined to produce the electromagnetic current,
\begin{equation}
V_\mu(x) = \frac{1}{2}V_\mu^u(x) - \frac{1}{2}V_\mu^d(x) \,.
\end{equation}

\begin{figure}
\includegraphics[width=10cm,clip=true]{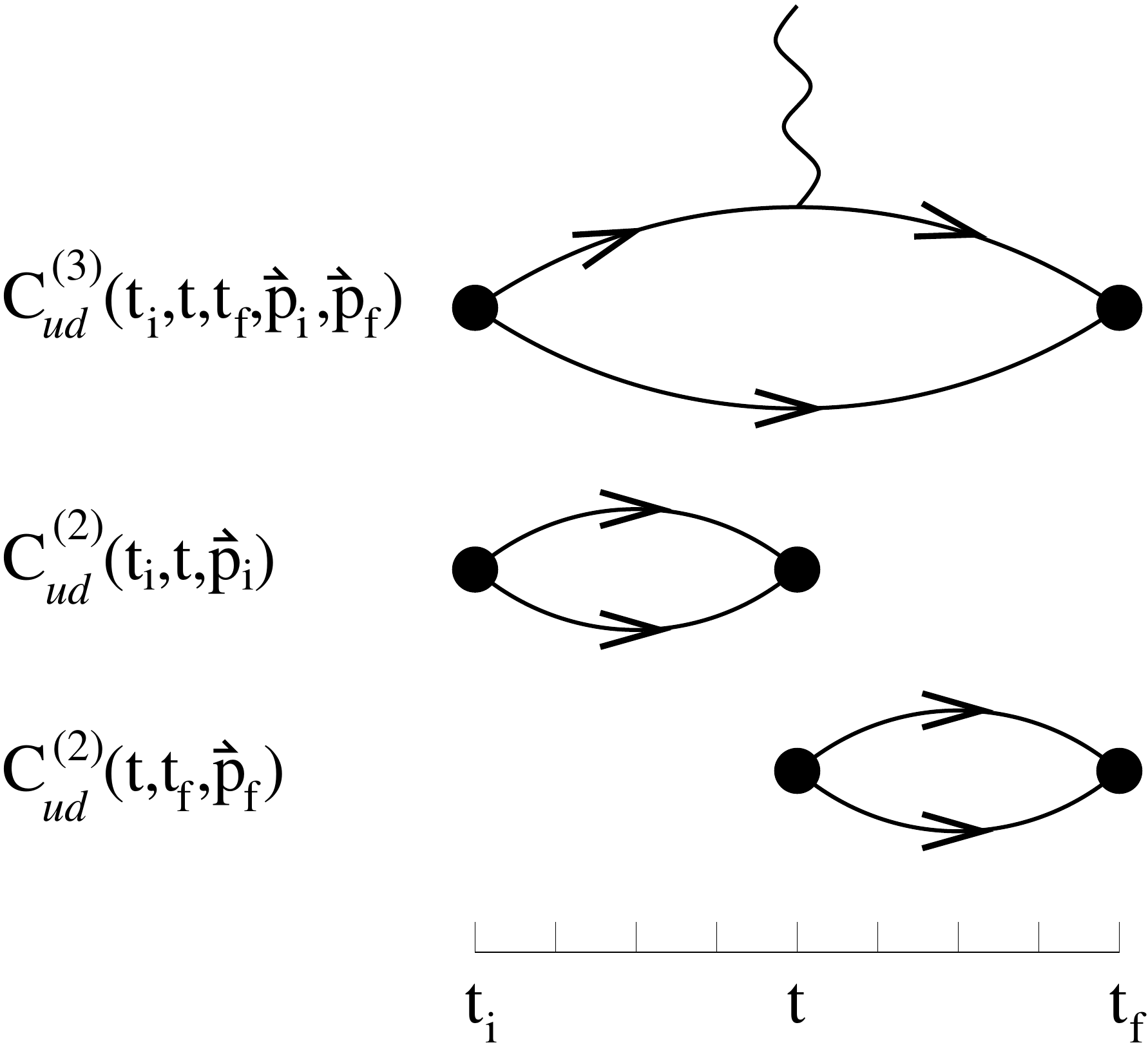}
\caption{The three correlation functions analyzed in a simultaneous fit to
         determine the mass and form factor of a Goldstone
         boson.  The central time $t$ is varied throughout the range
         $t_i<t<t_f$. The outgoing Goldstone boson momentum is chosen to be
         $\vec p_f=\vec 0$ in our simulations.}\label{fig:sketch}
\end{figure}

A three-point correlation function that probes the elastic form factor of the
DM candidate is
\begin{equation}
C^{(3)}_{ud}(t_i,t,t_f,\vec p_i,\vec p_f)
 = \sum_{\vec x_i,\vec x,\vec x_f}e^{-i(\vec x_f-\vec x)\cdot\vec p_f}
e^{-i(\vec x-\vec x_i)\cdot\vec p_i}
\left<0\left|{\cal O}_{ud}^{(\gamma_5)}(x_f)V_\mu(x)
{\cal O}_{ud}^{(\gamma_5)\dagger}(x_i)\right|0\right>
\end{equation}
where $\vec x$ denotes the spatial 3-vector within the 4-vector $x$.
A two-point correlation function represents particle propagation,
\begin{equation}
C^{(2)}_{ud}(t_i,t_f,\vec p)
 = \sum_{\vec x_i,\vec x_f}e^{-i(\vec x_f-\vec x_i)\cdot\vec p}
\left<0\left|{\cal O}_{ud}^{(\gamma_5)}(x_f)
{\cal O}_{ud}^{(\gamma_5)\dagger}(x_i)\right|0\right> \,.
\end{equation}
Two methods have been used for the lattice analysis, as has been done from the
earliest dynamical study of the pion form factor in SU(3) QCD \cite{Bonnet:2004fr}.
One method is to perform a simultaneous fit to the three correlation
functions shown pictorially in Fig.~\ref{fig:sketch}.
In particular, these correlation functions must be fit to their expected
hadronic forms:
\begin{eqnarray}
C^{(3)}_{ud}(t_i,t,t_f,\vec p_i,\vec p_f)
&=& \sum_{n_i}\sum_{n_f}Z_{n_f}\frac{e^{-(t_f-t)E_{n_f}(\vec p_f)}}
    {2E_{n_f}(\vec p_f)}\langle n_f(\vec p_f)|V_\mu(0)|n_i(\vec p_i)\rangle
    \frac{e^{-(t-t_i)E_{n_i}(\vec p_i)}}{2E_{n_i}(\vec p_i)}Z_{n_i}^* \,, \\
C^{(2)}_{ud}(t_i,t_f,\vec p)
&=& \sum_n|Z_n|^2\frac{e^{-(t_f-t_i)E_n(\vec p)}}{2E_n(\vec p)} \,.
\end{eqnarray}
In principle the sums include all hadrons having the quantum numbers of the
operator ${\cal O}_{ud}^{(\gamma_5)}$, but in practice only the lightest few
hadrons will be resolved by typical lattice data if $t_i$ and $t_f$ are
sufficiently far apart on the lattice.
In our simulations $C^{(3)}$ is dominated by the
ground state, i.e.\ the Goldstone boson of interest (generically named $\Pi$),
but excited states are
still observed in the pair of $C^{(2)}$ correlators.  Therefore we can fit to
\begin{eqnarray}
C^{(3)}_{ud}(t_i,t,t_f,\vec p_i,\vec p_f)
&=& |Z_\Pi|^2\frac{e^{-(t_f-t)E_\Pi(\vec p_f)}}{2E_\Pi(\vec p_f)}
    \frac{e^{-(t-t_i)E_\Pi(\vec p_i)}}{2E_\Pi(\vec p_i)}
    F_\Pi(Q^2)(p_i+p_f)_\mu \,, \\
C^{(2)}_{ud}(t_i,t_f,\vec p)
&=& |Z_\Pi|^2\frac{e^{-(t_f-t_i)E_\Pi(\vec p)}}{2E_\Pi(\vec p)}
  + \sum_{{\rm excited~}n}|Z_n|^2\frac{e^{-(t_f-t_i)E_n(\vec p)}}{2E_n(\vec p)}
  \,,
\end{eqnarray}
where we have used the standard definition of the form factor $F_\Pi(Q^2)$,
\begin{eqnarray}
\langle \Pi(\vec p_f)|V_\mu(0)|\Pi(\vec p_i)\rangle
    &=& F_\Pi(Q^2)(p_i+p_f)_\mu \,, \\
Q^2 &=& (\vec p_f-\vec p_i)^2 - \left(E_\Pi(\vec p_f)-E_\Pi(\vec p_i)\right)^2
        \,.
\end{eqnarray}
For any chosen lattice momentum, the fit parameters are the energies $E_\Pi$
and $E_n$, the coefficients $|Z_\Pi|^2$ and $|Z_n|^2$, and the form factor
$F_\Pi(Q^2)$.

Notice that our fitting functions are not periodic in the Euclidean time
direction.  Because a form factor calculation has three widely-spaced times,
$t_i$, $t$, and $t_f$, it is more economical to use a Dirichlet boundary
condition in the time direction for fermions.  Therefore the fitting functions
described above are the correct ones for our simulations.

The second method used for the lattice analysis, which gives results that are in
complete agreement with the first method, is known as the ratio method.
This second method uses an explicit formula for the form factor, valid
for $t_i\ll t \ll t_f$:
\begin{equation}\label{ratioeq}
F_\Pi(Q^2) = \frac{C_{UD}^{(3)}(t_i,t,t_f,\vec p_i,\vec p_f)C^{(2)}_{UD}(t_i,t,\vec p_f)}{C^{(2)}_{UD}(t_i,t,\vec p_i)C^{(2)}_{UD}(t_i,t_f,\vec p_f)}\left(\frac{2E_\Pi(\vec p_f)}{E_\Pi(\vec p_i)+E_\Pi(\vec p_f)}\right)
\end{equation}
It is straightforward to derive this expression from the preceding equations.
The ratio method is very convenient because all $Z_n$ have canceled away,
and the ratio $E_\Pi(\vec p_i)/E_\Pi(\vec p_f)$ is easy to obtain from the
lattice two-point functions.  All that remains is to fit the ratio to a
constant for each value of $Q^2$.
Another pleasant feature of Eq.~(\ref{ratioeq}) is that the only two-point
function that extends all the way from $t_i$ to $t_f$ has momentum $\vec p_f$.
Because we always choose $\vec p_f=\vec 0$, our simulations will provide
a precise numerical value for this factor in the ratio.

\section{Relationships among form factors}\label{sec:relations}

To determine what signal our DM candidate would induce to direct detection experiments, we estimate the electromagnetic form factors. The $u$ and $d$ fermions in our action have electroweak charges
that are constrained by anomaly cancellation: they form a left-handed weak
doublet, right-handed weak singlets, and have electric charges $Q_u=+1/2$ and
$Q_d=-1/2$.  Neither fermion carries QCD color.
The five Goldstone bosons have valence structure $\overline{u}d$,
$\overline{d}u$, $\frac{1}{\sqrt{2}}(\overline{u}u-\overline{d}d)$, $ud$ and
$\overline{u}\overline{d}$.  Because it is symmetric under $u\leftrightarrow d$,
the DM candidate $ud$ has no electroweak elastic form factors if there is no isospin breaking. Only the two electrically-charged Goldstones will have form factors in that
case. If a source of isospin breaking appears the electroweak elastic form factors will not vanish for
the DM candidate $ud$, and
they will be related to the form factors of the charged
Goldstones. Such a source of isospin breaking is naturally expected to
occur in Nature given that is already present for the ordinary quarks,
and moreover they are welcome because they can be used to further
diminish, or eliminate, the tension with the precision data. To mimic
this source of isospin breaking on the lattice we will simply assume two
different explicit masses for the up and down fermions.

Our lattice study can therefore follow the methods used for early
quenched studies of SU(2) gauge theory \cite{Wilcox:1985iy,Woloshyn:1985in,Woloshyn:1985vd}
and recent dynamical studies of SU(3) gauge theory
\cite{Bonnet:2004fr,Brommel:2006ww,Frezzotti:2008dr,Boyle:2008yd,Aoki:2009qn,Nguyen:2011ek,Brandt:2011jk,Brandt:2013dua},
with the difference that our fermions are dynamical. 

The five Goldstone bosons form a multiplet within the remaining Sp(4) global
symmetry, but that symmetry is not respected by electroweak interactions.
There are also deviations arising from $m_u\neq m_d$.
Here we derive some of the connections between correlation functions of the
Goldstone bosons.

To begin, we adapt a derivation provided in Ref.~\cite{Lewis:2011zb}
\begin{eqnarray}
C^{(2)}_{ud}(t_i,t_f,\vec p)
&=& \sum_{\vec x_i,\vec x_f}e^{-i(\vec x_f-\vec x_i)\cdot\vec p}
\left<0\left|{\cal O}_{ud}^{(\gamma_5)}(x_f)
{\cal O}_{ud}^{(\gamma_5)\dagger}(x_i)\right|0\right> \nonumber \\
&=& \sum_{\vec x_i,\vec x_f}e^{-i(\vec x_f-\vec x_i)\cdot\vec p}
    {\rm Tr}\left(u^T(x_f)(-i\sigma^2)C\gamma_5d(x_f)\overline{d}(x_i)
    \gamma_0\gamma_5^\dagger C^\dagger(-i\sigma^2)^\dagger\gamma_0^T
    \overline{u}^T(x_i)\right) \nonumber \\
&=& \sum_{\vec x_i,\vec x_f}e^{-i(\vec x_f-\vec x_i)\cdot\vec p}
    {\rm Tr}\left(\overline{u}(x_f)\gamma_5d(x_f)
    \overline{d}(x_i)\gamma_0\gamma_5^\dagger\gamma_0u(x_i)\right) \nonumber \\
&=& C^{(2)}_{\overline{u}d}(t_i,t_f,\vec p)
\end{eqnarray}
where we have made use of two properties of the charge conjugation operator:
\begin{eqnarray}
\gamma^{\mu T} &=& -C\gamma^\mu C^\dagger \,, \\
\left[u(y)\overline{u}(x)\right]^T
&=& C(-i\sigma^2)u(x)\overline{u}(y)C^\dagger(-i\sigma^2)^\dagger \,.
\end{eqnarray}
Similar derivations lead to the following relations among three-point
correlation functions,
\begin{eqnarray}
C^{(3)}_{ud}(t_i,t,t_f,\vec p_i,\vec p_f) &=& T^u - T^d \,, \\
C^{(3)}_{\overline{ud}}(t_i,t,t_f,\vec p_i,\vec p_f) &=& -T^u + T^d \,, \\
C^{(3)}_{u\overline{d}}(t_i,t,t_f,\vec p_i,\vec p_f) &=& T^u + T^d \,, \\
C^{(3)}_{\overline{u}d}(t_i,t,t_f,\vec p_i,\vec p_f) &=& -T^u - T^d \,, \\
C^{(3)}_{\overline{u}u+\overline{d}d}(t_i,t,t_f,\vec p_i,\vec p_f) &=& 0 \,,
\end{eqnarray}
where
\begin{equation}
T^X = \sum_{\vec x_i,\vec x,\vec x_f}e^{-i(\vec x_f-\vec x)\cdot\vec p_f}
      e^{-i(\vec x-\vec x_i)\cdot\vec p_i}
      \left<0\left|{\cal O}_{ud}^{(\gamma_5)}(x_f)V^X_\mu(x)
      {\cal O}_{ud}^{(\gamma_5)\dagger}(x_i)\right|0\right> \,.
\end{equation}
For the special case of $m_u=m_d$, we find $T^u=T^d$ so only the charged
Goldstones, $\overline{u}d$ and $\overline{d}u$, have a nonzero form factor.
In the general case of $m_u\neq m_d$, we see that the DM candidate
$ud$ (and its antiparticle) also has a form factor.

Lattice simulations could in principle determine $T^u$ and $T^d$ in the
general case, but they contain contributions from quark-disconnected diagrams
that would require
significant computational resources.  Lattice simulations with $m_u=m_d$ are
more manageable, but then $T^u=T^d$ so there is no DM form factor in
that case.

There is an explicit relationship between $T^u$ and $T^d$ in the large $N_c$
limit.  In that limit hadronic resonances become narrow, so $T^u$ and $T^d$
are each written as a sum over vector meson
poles\cite{'tHooft:1973jz,Witten:1979kh,Masjuan:2012sk}.
In practice those sums are dominated by the lightest vector mesons.
Perhaps surprisingly, this large $N_c$ result has long been known to work
rather well for QCD despite the seemingly small value of $N_c=3$.
For example, the $\pi^+$ form factor is
dominated by $\rho^0$ meson exchange and the $K^+$ form factor is dominated by
$\rho^0$ and $\phi$ meson exchange,
\begin{eqnarray}
F_{\pi^+}(Q^2) &\approx& \frac{2}{3}\left(\frac{m_\rho^2}{m_\rho^2+Q^2}\right)
                       + \frac{1}{3}\left(\frac{m_\rho^2}{m_\rho^2+Q^2}\right)
         \,, \\
F_{K^+}(Q^2) &\approx& \frac{2}{3}\left(\frac{m_\rho^2}{m_\rho^2+Q^2}\right)
                     + \frac{1}{3}\left(\frac{m_\phi^2}{m_\phi^2+Q^2}\right)
         \,.
\end{eqnarray}
QCD also contains an example that
exactly parallels our $m_u\neq m_d$ effects:
the neutral kaon has a nonzero form factor arising from $m_d\neq m_s$.
The experimental determination of the neutral kaon charge
radius\cite{Beringer:1900zz} is dominated by the
difference between $\rho^0$ and $\phi$ meson exchanges,
\begin{eqnarray}
F_{K^0}(Q^2) &\approx& -\frac{1}{3}\left(\frac{m_\rho^2}{m_\rho^2+Q^2}\right)
                 + \frac{1}{3}\left(\frac{m_\phi^2}{m_\phi^2+Q^2}\right) \,, \\
\langle r^2\rangle_{K^0} &=& \left.-6\frac{dF_{K^0}}{dQ^2}\right|_{Q^2=0} \,.
\end{eqnarray}

If the large $N_c$ result were also applicable to our $N_c=2$ technicolor
template, then lattice determinations of the vector meson masses would
provide estimates of all Goldstone form factors.  Moreover, the dark
matter form factors would be related to $W^\pm$ form factors.
In the following section we will perform a lattice simulation of the
Goldstone form factor in the $m_u=m_d$ limit, i.e.\ $T^u+T^d$, and show
that the large $N_c$ result does indeed hold to a good accuracy in our $N_c=2$ theory.

\section{The Lattice Results}\label{sec:latresults}

The numerical work in this paper is based on the same configurations generated
 in \cite{Hietanen:2014xca}.
A complete analysis of 500 configurations at $(\beta,m_0)=(2.2,-0.72)$
provides a first result for the form factor.  To consider discretization
effects an analysis of 300 configurations at $(\beta,m_0)=(2.0,-0.947)$ is
performed.  To study chiral extrapolation effects, an analysis of 300
configurations at $(\beta,m_0)=(2.2,-0.75)$ is performed.
All ensembles were created with the HiRep code \cite{DelDebbio:2008zf} for
fully-dynamical
plaquette-action SU(2) gauge theory with two flavors of
mass-degenerate Wilson fermions.

\begin{figure}
  \includegraphics[width=16cm,clip=true,trim=30 40 30 80]{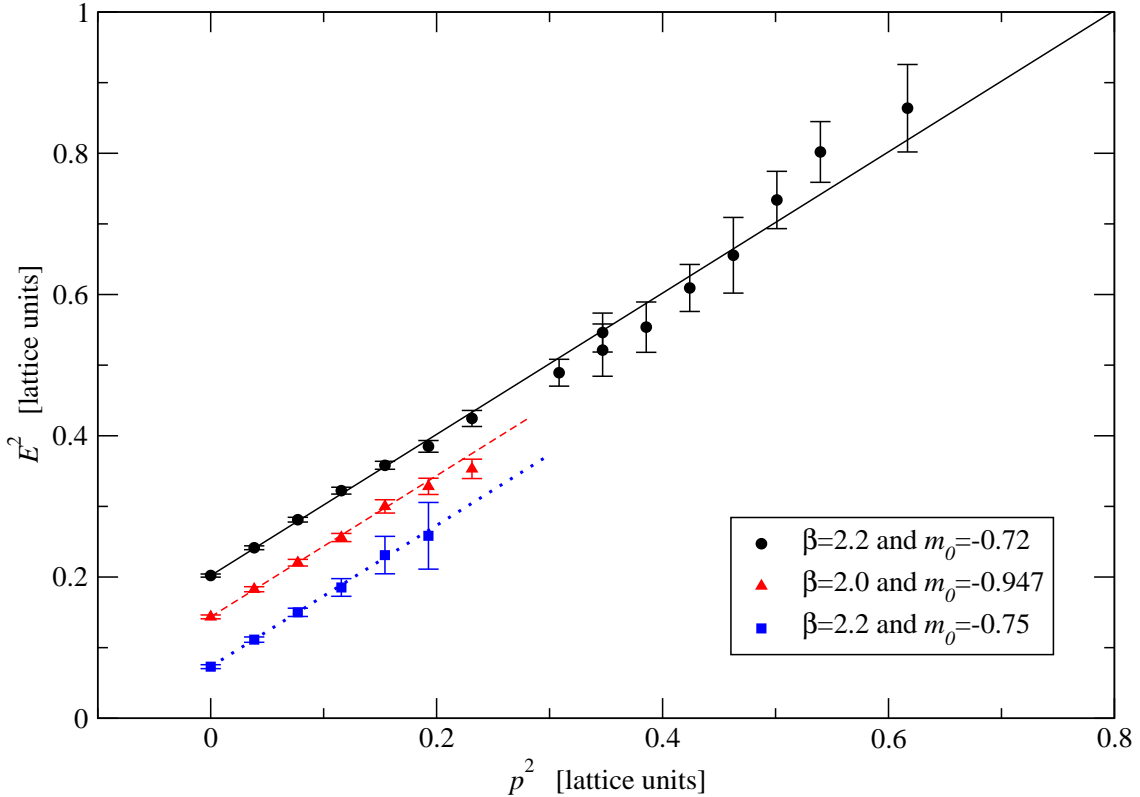}
  \caption{Squared energies of Goldstone bosons as functions of
    $p^2=p_x^2+p_y^2+p_z^2$.
    Straight lines are $m_\Pi^2+p^2$ for the measured lattice mass $m_\Pi$.
  }\label{fig:pseudoscalar}
\end{figure}

Extraction of the form factor requires the energies of Goldstone bosons
that are moving across the lattice.  There is a direct relationship in the
continuum,
\begin{equation}\label{continuumdispersion}
E^2 = m^2+p^2 \,,
\end{equation}
and also on the lattice
\begin{eqnarray}
\hat E^2 &=& 4\sinh^2(m/2) + \hat p^2 \,, \label{latticedispersion} \\
\hat E &\equiv& 2\sinh(E/2) \,, \\
\hat p^2 &\equiv& 4\sum_{i=1}^3\sin^2(p_i/2) \,.
\end{eqnarray}
Figure~\ref{fig:pseudoscalar} shows three straight lines that represent the
continuum relation; the only input for those lines is the Goldstone mass because their slopes are completely determined by kinematics.  Direct lattice
computations of the energy of a moving Goldstone boson are also shown.
Note that lattice discretization provides access to
\begin{equation}\label{latmom}
\vec p = \frac{2\pi}{L}\left(k_x\hat x+k_y\hat y+k_z\hat z\right)
\end{equation}
where $L=32$ and we use $0\leq k_i\leq 3$.
Since the data presented in the Figure~\ref{fig:pseudoscalar} lie on the
continuum lines, there is no indication of any discretization errors.
This conclusion is true of all three data sets up to $p^2\sim0.6$, though only
one data set was shown for the full range to avoid cluttering the plot.

\begin{figure}
\includegraphics[width=16cm,clip=true,trim=30 40 30 80]{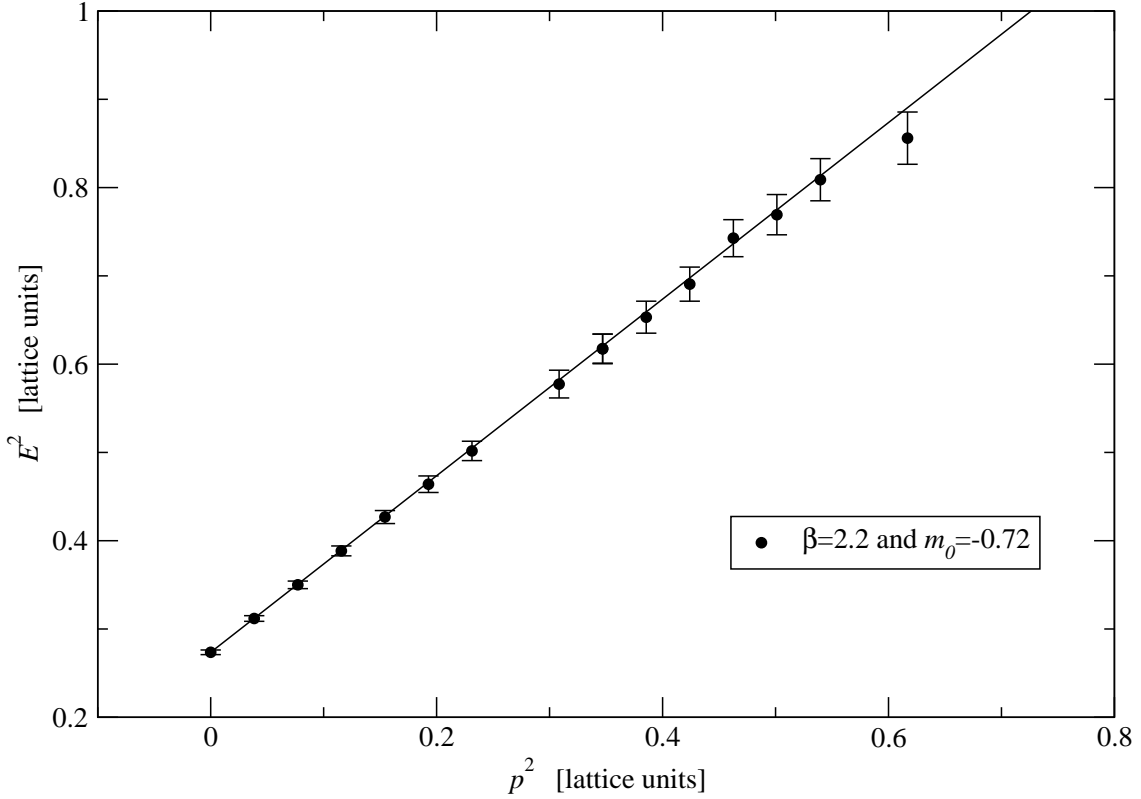}
\caption{Squared energy of the vector meson as a function of
         $p^2=p_x^2+p_y^2+p_z^2$, for $(\beta,m_0)=(2.2,-0.72)$.
         The straight line is $m_V^2+p^2$ for the measured lattice mass $m_V$.
         }\label{fig:vector}
\end{figure}

\begin{figure}
\includegraphics[width=16cm,clip=true,trim=30 40 30 80]{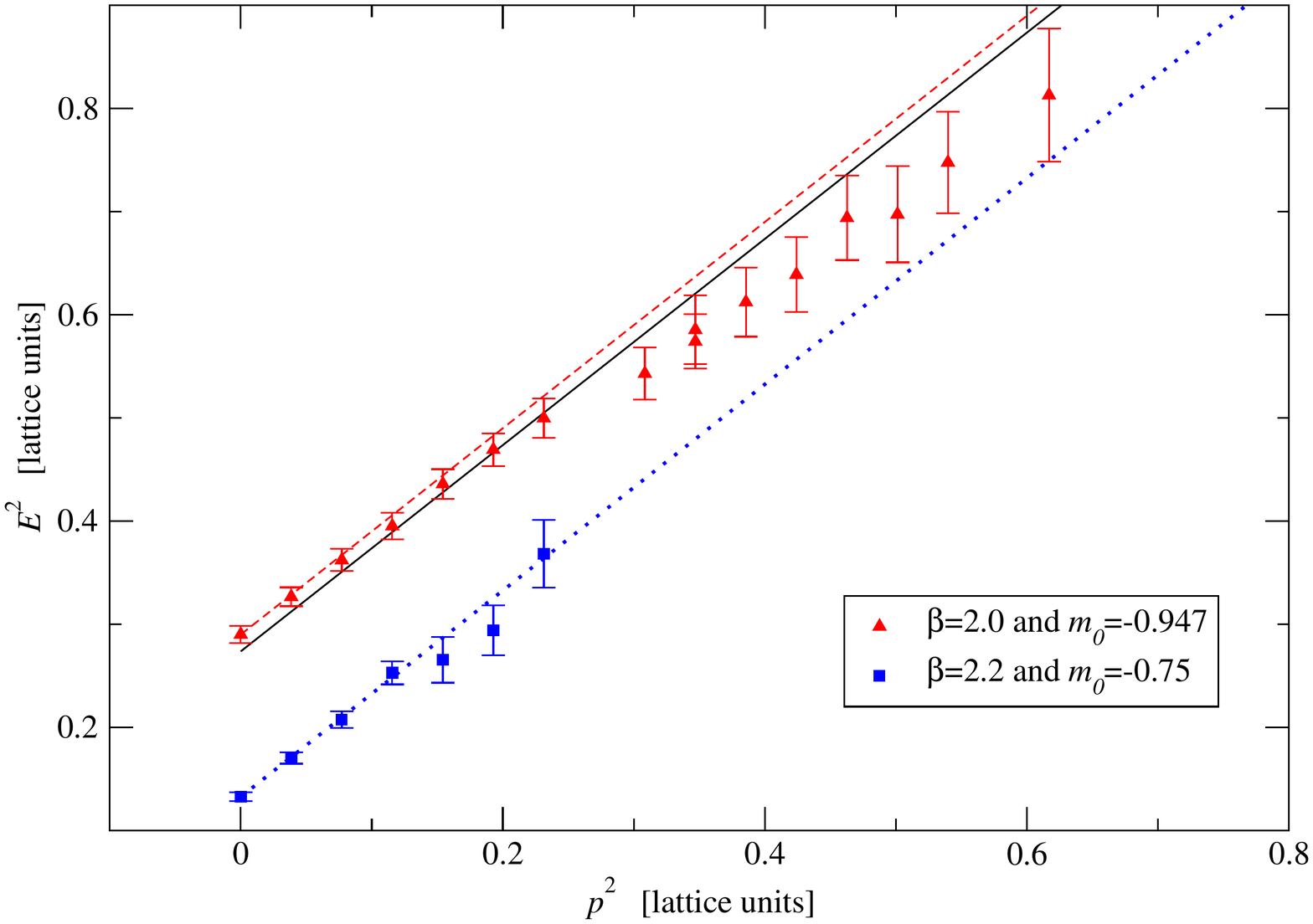}
\caption{Squared energies of the vector meson as functions of
         $p^2=p_x^2+p_y^2+p_z^2$, for $(\beta,m_0)=(2.0,-0.947)$ and
         $(\beta,m_0)=(2.2,-0.75)$.
         Straight lines are $m_V^2+p^2$ for the measured lattice mass $m_V$:
         dashed line for $(\beta,m_0)=(2.0,-0.947)$,
         dotted line for $(\beta,m_0)=(2.2,-0.75)$, and (for comparison)
         solid line for $(\beta,m_0)=(2.2,-0.72)$.
         }\label{fig:vector2}
\end{figure}

\begin{figure}
\includegraphics[width=16cm,clip=true,trim=30 40 30 80]{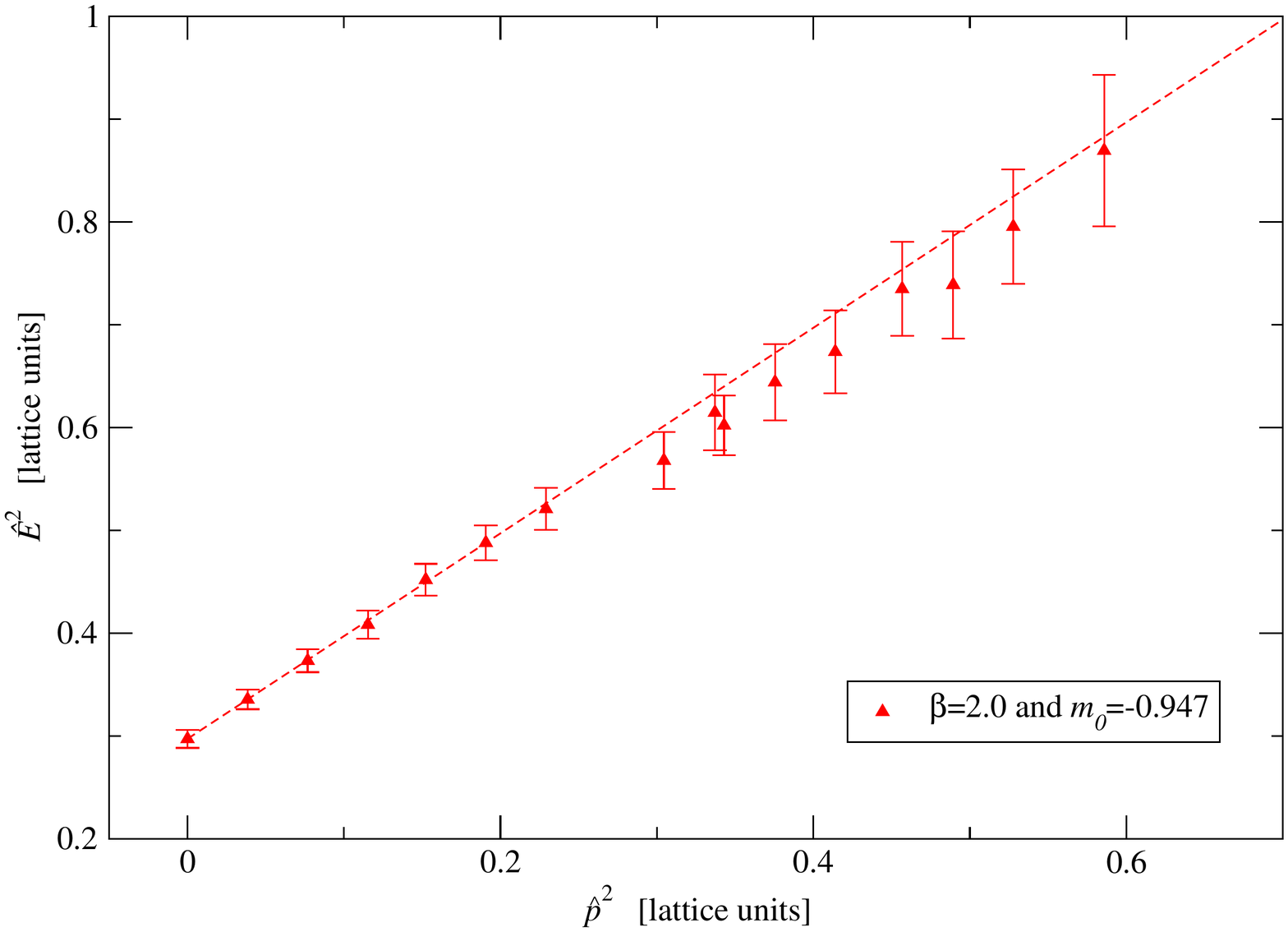}
\caption{Squared energy of the vector meson as a function of
         $\hat p^2=4\sin^2(p_x/2)+4\sin^2(p_y/2)+4\sin^2(p_z/2)$,
         for $(\beta,m_0)=(2.0,-0.947)$.
         The straight line is $4\sinh^2(m_V/2)+\hat p^2$ for the measured
         lattice mass $m_V$.
         }\label{fig:latdispersion}
\end{figure}

More precise lattice data are obtained for the vector meson, in part
because of the ability to average over all three polarizations.
Figure~\ref{fig:vector} shows the agreement with continuum expectations
for one data set.  The mass agrees with Ref.~\cite{Lewis:2011zb} and the
momentum dependence agrees with the continuum line.
The other two data sets are displayed in Fig.~\ref{fig:vector2} with the
coarser lattice extended as far as $p^2\approx0.6$.
For comparison, that same data set is compared to the lattice expectation
in Fig.~\ref{fig:latdispersion} where the required agreement is seen for all
momentum values.
Though discretization effects are modest, we will ensure self-consistency
by using lattice relations
rather than continuum relations when analyzing the form factor.

The vector meson is of interest to the present work because the Goldstone
boson form factor is expected to exhibit vector meson dominance.
The straight lines in Fig.~\ref{fig:vector2} indicate that two of our
ensembles have nearly-equal vector meson masses in lattice units, suggesting
that their form factors should also be similar, although
Fig.~\ref{fig:pseudoscalar} shows that their Goldstone masses are not equal.

We choose the outgoing
Goldstone to be at rest in our form factor computations,
so momentum flows from the incoming Goldstone to the
photon coupling.  All momentum directions are averaged for each configuration;
for example, form factors with
$(k_x,k_y,k_z)=(1,0,2)$, (1,2,0), (2,0,1), (2,1,0), (0,1,2) and
(0,2,1) in Eq.~(\ref{latmom})
are all computed and averaged to help reduce statistical errors.
We use Dirichlet boundary conditions in the time direction for
fermions in the measurements,
meaning that fermions do not propagate beyond the lattice's temporal
boundaries. However, the configurations were generated using periodic
boundary conditions in time directions. The Goldstone
creation operator is placed at the fifth time step from the 
lattice's left edge ($t_i=4$) and the annihilation operator is placed at
the fifth from the right ($t_f=27$).

\begin{figure}
\includegraphics[width=16cm,clip=true,trim=30 40 30 80]{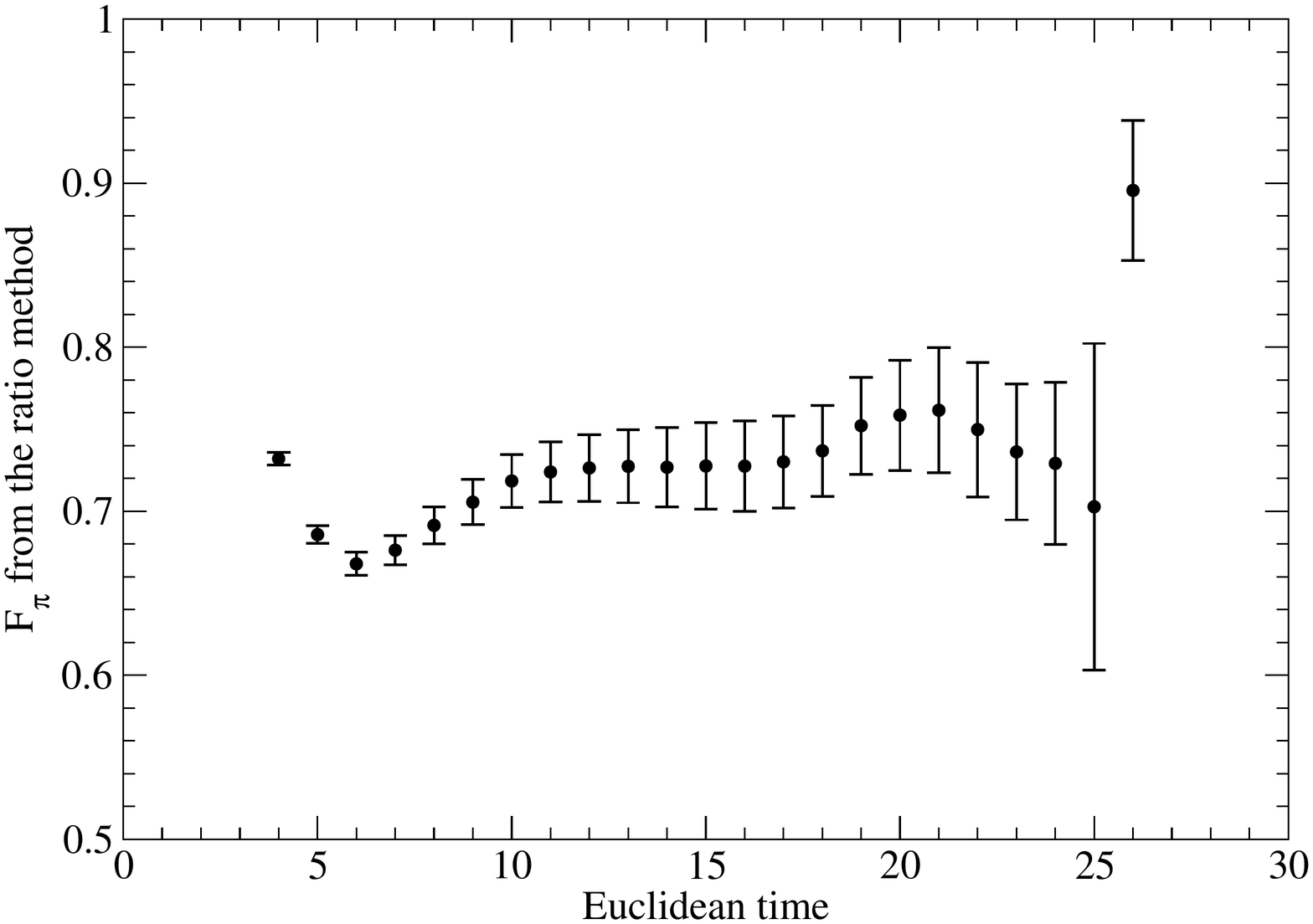}
\caption{The ratio definition of the Goldstone boson form factor,
         Eq.~(\ref{ratioeq}), for momentum $(k_x,k_y,k_z)=(1,1,0)$ in the
         ensemble having $(\beta,m_0)=(2.2,-0.72)$.
         The energetic Goldstone is created at $t_i=5$, the stationary
         Goldstone is annihilated at $t_f=27$, and the ratio should be fit to
         a constant for Euclidean times $t$ that satisfy $t_i\ll t\ll t_f$.
         }\label{fig:ratio}
\end{figure}
As an example of lattice data for the form factor, Fig.~\ref{fig:ratio}
shows the raw form factor data for the right-hand side of Eq.~(\ref{ratioeq})
with one particular momentum in the $(\beta,m_0)=(2.2,-0.72)$ ensemble.
There is a broad range of Euclidean times between $t_i$ and $t_f$ where the
ratio is indeed constant, allowing the form factor to be read from the plot.
When a similar plot is made for vanishing momentum, the form factor is
{\em exactly} equal to unity due to our use of the conserved vector current
which obeys the corresponding lattice Ward-Takahashi identity.

The four-momentum transfer is defined by
\begin{equation}
q = (\vec p_f-\vec p_i,E_f-E_i)
\end{equation}
and putting that into the continuum dispersion relation, Eq.~(\ref{continuumdispersion}), gives
\begin{equation}
Q^2 \equiv -q^2 = (\vec p_f-\vec p_i)^2 - (E_f-E_i)^2
\end{equation}
while putting it into the lattice dispersion relation, Eq.~(\ref{latticedispersion}), gives
\begin{equation}
\hat Q^2 \equiv -q^2 = -4{\rm arcsinh}^2\sqrt{\sinh^2\left(\frac{E_f-E_i}{2}\right)-\sum_{j=x,y,z}\sin^2\left(\frac{(p_f-p_i)_j}{2}\right)}
\end{equation}
Any difference between the lattice and continuum expressions is due to discretization errors that are small for our ensembles.

\begin{figure}
\includegraphics[width=16cm,clip=true,trim=30 40 30 80]{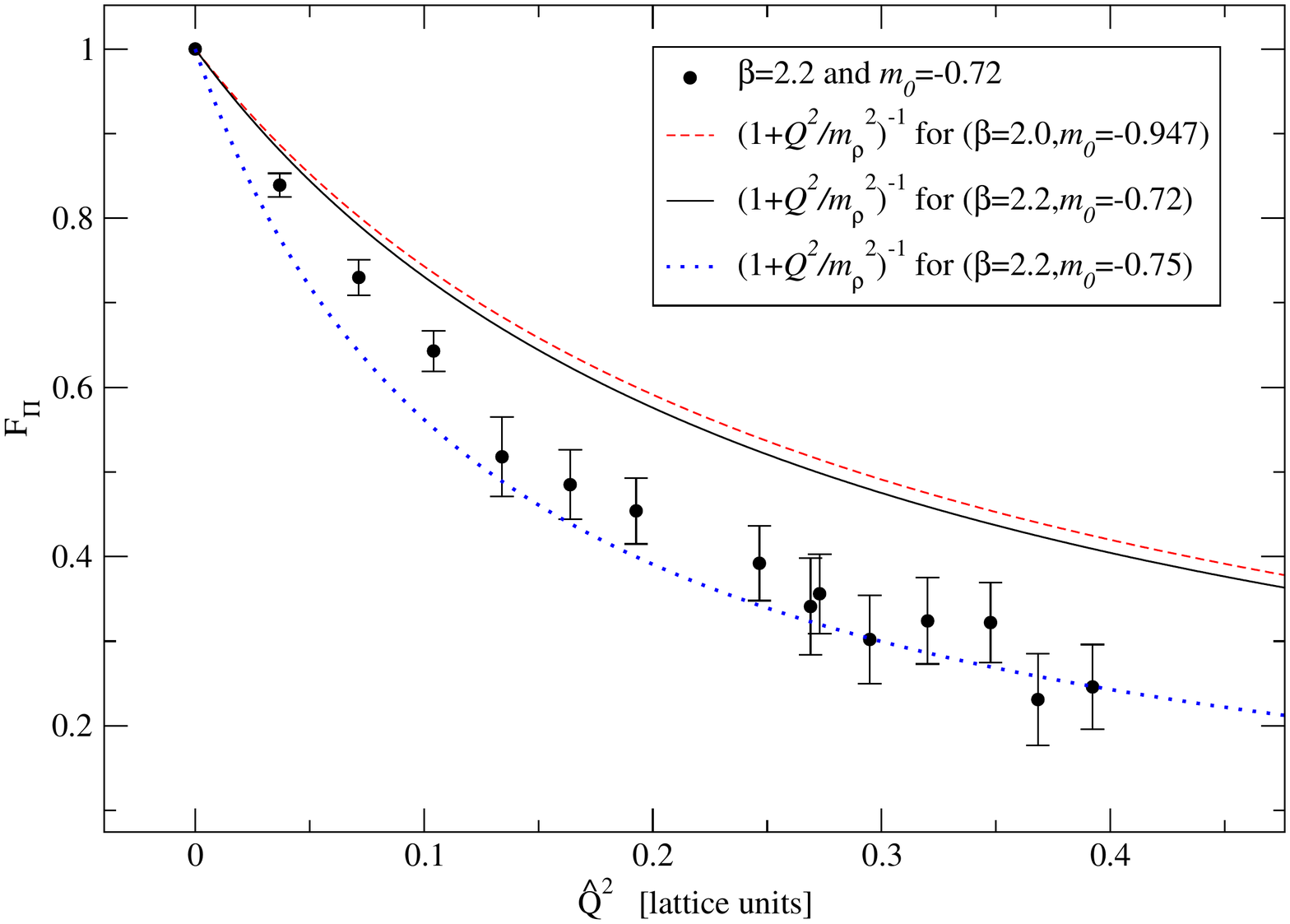}
\caption{Lattice result for the Goldstone form factor at $(\beta,m_0)=(2.2,-0.72)$.
         The solid curve is the prediction from a simple vector meson pole
         with vector mass taken directly from our lattice simulation.
         The dashed and dotted curves are shown only to aid comparison with
         Figs.~\ref{fig:ff2} and \ref{fig:ff3}.
         }\label{fig:ff1}
\end{figure}
\begin{figure}
\includegraphics[width=16cm,clip=true,trim=30 40 30 80]{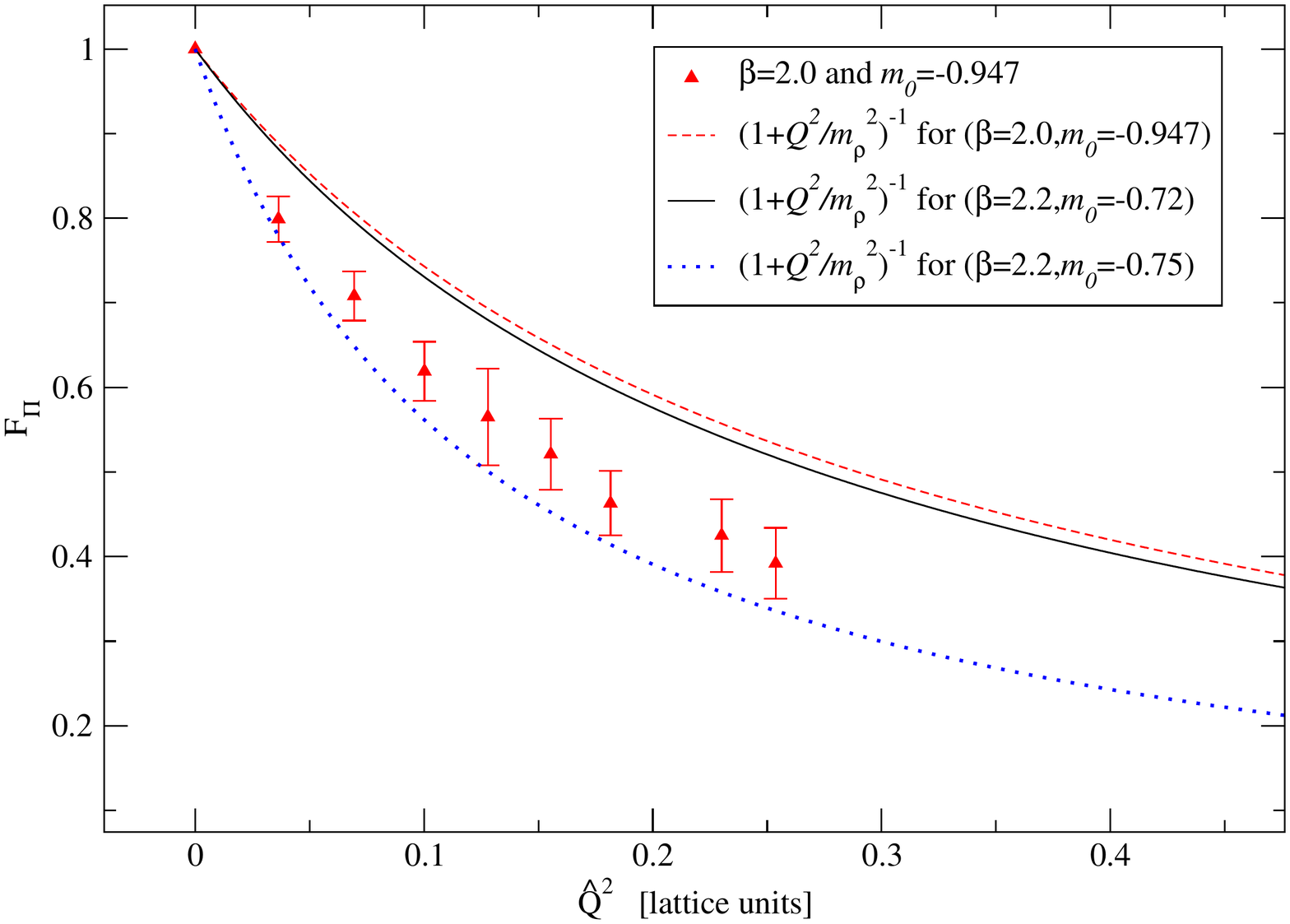}
\caption{Lattice result for the Goldstone form factor at $(\beta,m_0)=(2.0,-0.947)$.
         The dashed curve is the prediction from a simple vector meson pole
         with vector mass taken directly from our lattice simulation.
         The solid and dotted curves are shown only to aid comparison with
         Figs.~\ref{fig:ff1} and \ref{fig:ff3}.
         }\label{fig:ff2}
\end{figure}

Numerical results for the form factor at $(\beta,m_0)=(2.2,-0.72)$ are shown
in Fig.~\ref{fig:ff1}.  The lattice data have the shape
of a simple vector meson pole, but with a mass parameter significantly
different from the lattice vector meson mass.
As mentioned previously, our coarser lattice has almost the same vector meson
mass so it should give essentially the same form factor, and Fig.~\ref{fig:ff2}
verifies this expectation.  It too is thus significantly below its vector
meson pole.

\begin{figure}
\includegraphics[width=16cm,clip=true,trim=30 40 30 80]{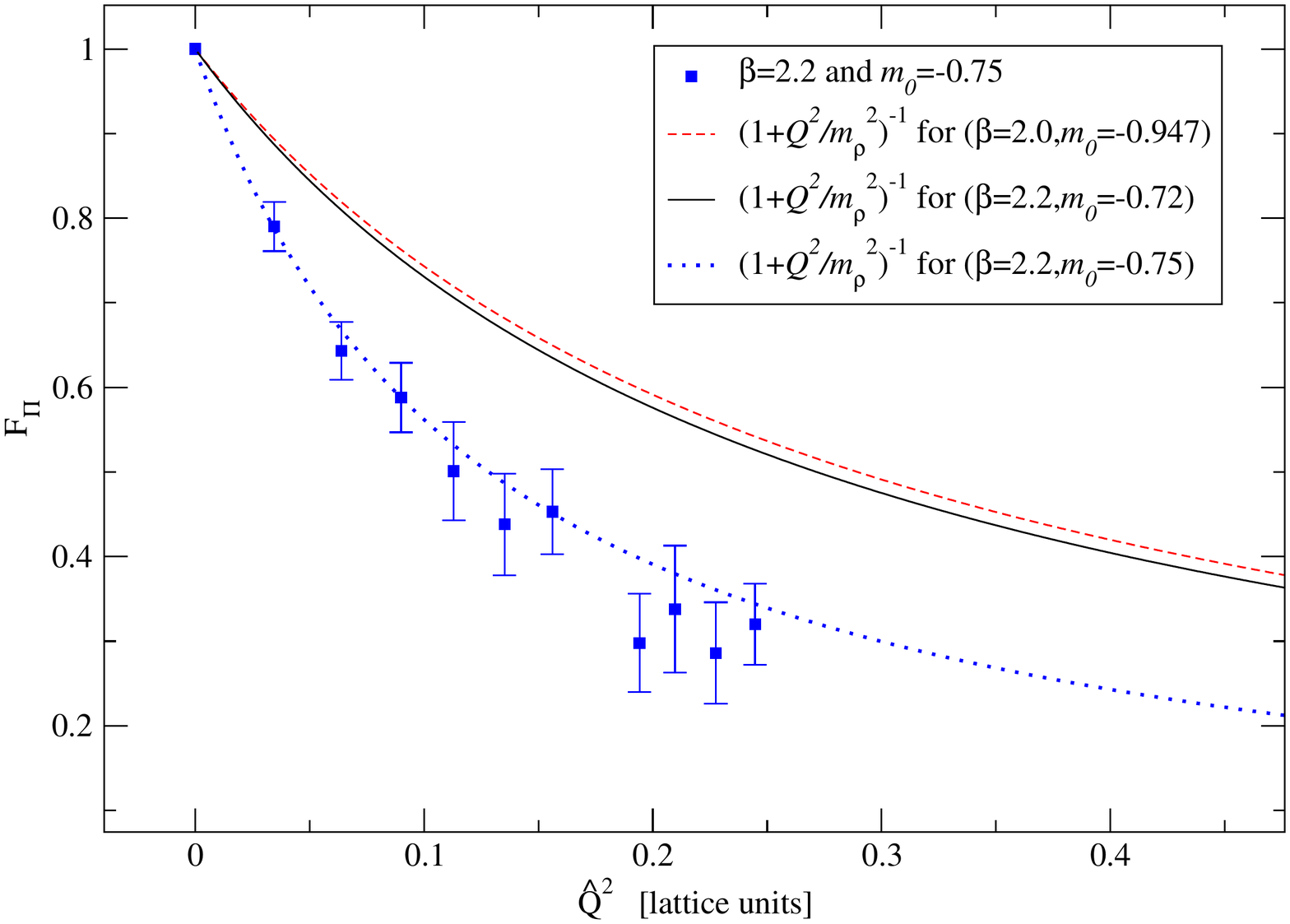}
\caption{Lattice result for the Goldstone form factor at $(\beta,m_0)=(2.2,-0.75)$.
         The dotted curve is the prediction from a simple vector meson pole
         with vector mass taken directly from our lattice simulation.
         The solid and dashed curves are shown only to aid comparison with
         Figs.~\ref{fig:ff1} and \ref{fig:ff2}.
         }\label{fig:ff3}
\end{figure}

It is no surprise that the lightest vector meson does not explain the entire
Goldstone form factor.  For QCD, chiral perturbation theory contains
correction terms suppressed by powers of $m_\pi^2/\Lambda_\chi^2$ \cite{Gasser:1984gg} and
similar terms are present in our SU(2) theory.  In fact, the SU(2) theory
has five Goldstone bosons instead of only three.  If such chiral terms are
responsible for the difference between lattice results and the vector pole,
then that difference should be reduced when the fermion mass is reduced.
Figure~\ref{fig:ff3} supports this view by showing that the vector meson
pole is in statistical agreement with lattice results at our lightest fermion
mass.

\section{Photon - Dark matter form factor: the basics}\label{sec:phenoresults}

To make predictions for experimental searches, we can follow the
general framework for technicolor DM developed in \cite{Foadi:2008qv}.
A more elaborate discussion is presented in following sections.  Here we
give only some basics.

The charge radius of a scalar couples to the photon as follows:
\begin{equation}
\mathcal{L}_B=i e \frac{d_B}{\Lambda^2} \phi^*\overleftrightarrow{\partial_\mu} \phi \, \partial_{\nu}F^{\mu\nu}
\end{equation}
In our case we have a specific expression for the coefficient,
\begin{eqnarray}
\frac{d_B}{\Lambda^2}
&=& \lim_{Q^2\to0}\frac{1}{Q^2}\left[\frac{1}{2}\frac{m_{\rho_u}^2}{m_{\rho_u}^2+Q^2}
  - \frac{1}{2}\frac{m_{\rho_d}^2}{m_{\rho_d}^2+Q^2}\right] \\
&=& \frac{m_{\rho_u}^2-m_{\rho_d}^2}{2m_{\rho_u}^2m_{\rho_d}^2}
\end{eqnarray}
which, for small isospin breaking ($m_{\rho_u}\approx m_{\rho_d}\equiv m_\rho$),
corresponds to
\begin{eqnarray}
\Lambda &=& m_\rho \,, \\
d_B &=& (m_{\rho_u}-m_{\rho_d})/m_\rho \,.
\end{eqnarray}
For the numerical value of $m_\rho$ we use $2.5\pm 0.5 {\rm ~TeV}$
\cite{Hietanen:2014xca}.


Also from \cite{Foadi:2008qv}, the cross section for a DM particle
$\phi$ scattering from a nucleon through photon exchange is
\begin{equation}
\sigma_p^\gamma = \frac{\mu^2}{4\pi}\left(\frac{8\pi\alpha d_B}{\Lambda^2}
                  \right)^2
\end{equation}
where $\mu=m_\phi m_N/(m_\phi+m_N)$.
Assuming $m_\phi>m_N$, we see that $m_N/2\mu<m_N$ and the only remaining
unknown is $|d_B|$ which is clearly less than unity. 
We therefore have an upper bound on the cross section in this
model, \footnote{Note that we could perform the simulations with
  degenerated fermion masses as the isospin breaking is only parameterized by the small
unknown $d_B$.}
\begin{equation}
\sigma_p^\gamma < 2.3\times10^{-44} {\rm ~cm}^2 \,.
\end{equation}
However, it is important to consider the cross section
for scattering through Higgs exchange as well, which can interfere
with photon exchange. This issue will be addressed in the upcoming 
Sec.~\ref{sec:experiments}.

 \section{Adding the composite Higgs}\label{sec:experiments}
\vspace{-.2cm}

Besides the photon  interactions we expect also a composite Higgs exchange \cite{Foadi:2008qv,Ryttov:2008xe,Belyaev:2010kp,DelNobile:2011je}.  The relevant, for detection experiments, Lagrangian terms between our DM candidate and the composite Higgs are
 \begin{eqnarray}
   \frac{d_1}{\Lambda}  h \, \partial_{\mu}\phi^{\ast}\partial^{\mu}\phi +  \frac{d_2}{\Lambda} m^2_{\phi}\, h \,\phi^{\ast}\phi   \ . 
\end{eqnarray}
We have taken into account the pseudo-Goldstone nature of the DM field $\phi$ and therefore we expect $d_1$ and $d_2$ to be order unity.  

Making the further minimal assumption that the composite Higgs state couples to the standard model fermions with a strength proportional to their masses, as it is for the ordinary Higgs, the zero momentum transfer cross section of $\phi$  scattering off a nucleus with $Z$ protons and $A - Z$ neutrons is \cite{Foadi:2008qv, Belyaev:2010kp}
\begin{equation}\label{sigma_A}
\sigma_A = \frac{\mu_A^2}{4 \pi} \left| Z f_p + (A - Z) f_n \right|^2 \ ,
\end{equation}
where
\beq
f_n = d_H f \frac{m_p}{m_H^2 m_\phi} \ , \qquad\qquad f_p = f_n - \frac{8 \pi \alpha d_B}{\Lambda^2} \ , \label{fnfp}
\eeq
$m_p$ is the nucleon mass, $\mu_A$ is the $\phi$-nucleus reduced mass and~$f \sim 0.3$~parametrizes the Higgs to nucleon coupling and  we have defined \cite{DelNobile:2011je}: 
\begin{equation}
d_H =   -\frac{d_1 + d_2}{v_{EW}\,\Lambda} m^2_{\phi} \ .
\end{equation}

The event rate for generic couplings $f_n$ and $f_p$ is 
\beq\label{R}
R = \sigma_p \sum_i \eta_i \frac{\mu_{A_i}^2}{\mu_p^2} I_{A_i} \left| Z + (A_i - Z) f_n / f_p \right|^2 \ ,
\eeq
where $\eta_i$ is the abundance of the specific isotope $A_i$ in the detector material, and $I_{A_i}$ contains all the astrophysical factors as well as  the nucleon form factor $ F_{A_i} (E_R)$. For a given isotope we have
\beq
I_{A_i} = N_T \, n_\phi \int \ud E_R \int_{v_\text{min}}^{v_\text{esc}} \ud^3 v \, f(v) \frac{m_{A_i}}{2 v \mu_{A_i}^2} F_{A_i}^2 (E_R) \ .
\eeq
Here $m_{A_i}$ is the mass of the target nucleus, $N_T$ is the number of target nuclei, $n_\phi$ is the local number density of DM particles, and $f(v)$ is their local velocity distribution. The velocity integration is limited between the minimum velocity required in order to transfer a recoil energy $E_R$ to the scattered nucleus, $v_\text{min} = \sqrt{m_A E_\text{R} / 2 \mu_A^2}$, 
and the escape velocity from the galaxy $v_\text{esc}$. The $\phi$-proton cross section $\sigma_p =  {\mu_p^2} \left| f_p \right|^2 /{4 \pi}$ can be easily obtained by setting $A = Z = 1$ in Eq.~\eqref{sigma_A}.
 
Direct DM search collaborations quote constraints on generic WIMP-nuclei cross sections normalized to the WIMP-nucleon cross section $\sigma_p^\text{exp}$ (assuming conventionally  $f_n = f_p$).  Therefore the experimentally constrained event rate can be cast in the following form
\beq \label{Rexp}
R = \sigma_p^\text{exp} \sum_i \eta_i \frac{\mu_{A_i}^2}{\mu_p^2} I_{A_i} A_i^2 \ .
\eeq
  Equating Eqs.~\eqref{R} and \eqref{Rexp} yields the experimental constraints on the generic WIMP-proton cross section $\sigma_p$ with arbitrary couplings $f_p$ and $f_n$
\beq\label{sigma_pTOT}
\sigma_p = \sigma_p^\text{exp} \frac{\sum_i \eta_i \mu_{A_i}^2 I_{A_i} A_i^2}{\sum_i \eta_i \mu_{A_i}^2 I_{A_i} \left| Z + (A_i - Z) f_n / f_p \right|^2} \ .
\eeq
Provided that the factors $I_{A_i}$ do not change significantly from one  isotope to another as it is the case \cite{DelNobile:2011je}, they drop out from the ratio.

\begin{figure}[h!]
\includegraphics[width=.5\textwidth
]{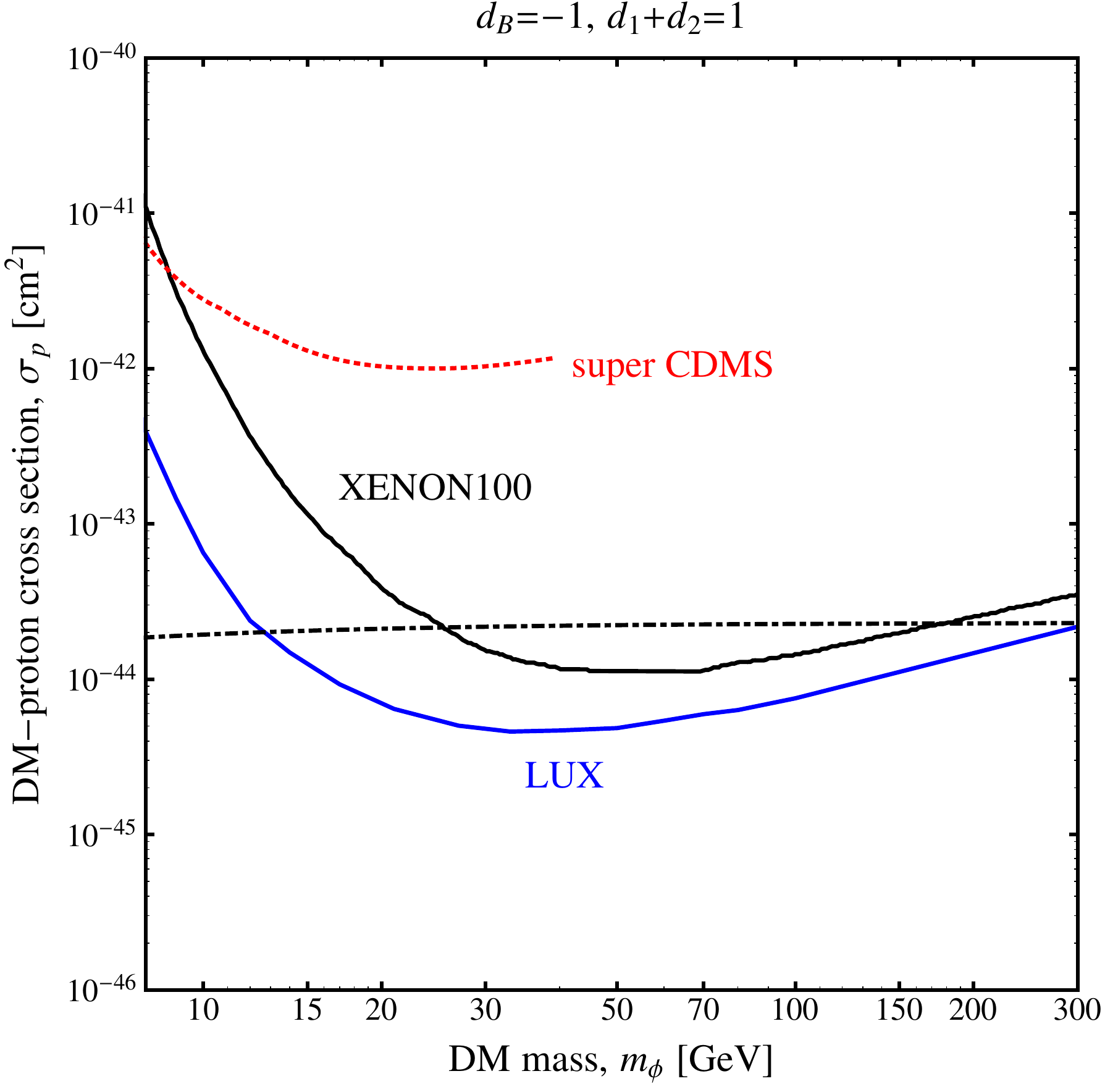}~
\vskip .6cm
\includegraphics[width=.5
\textwidth
]{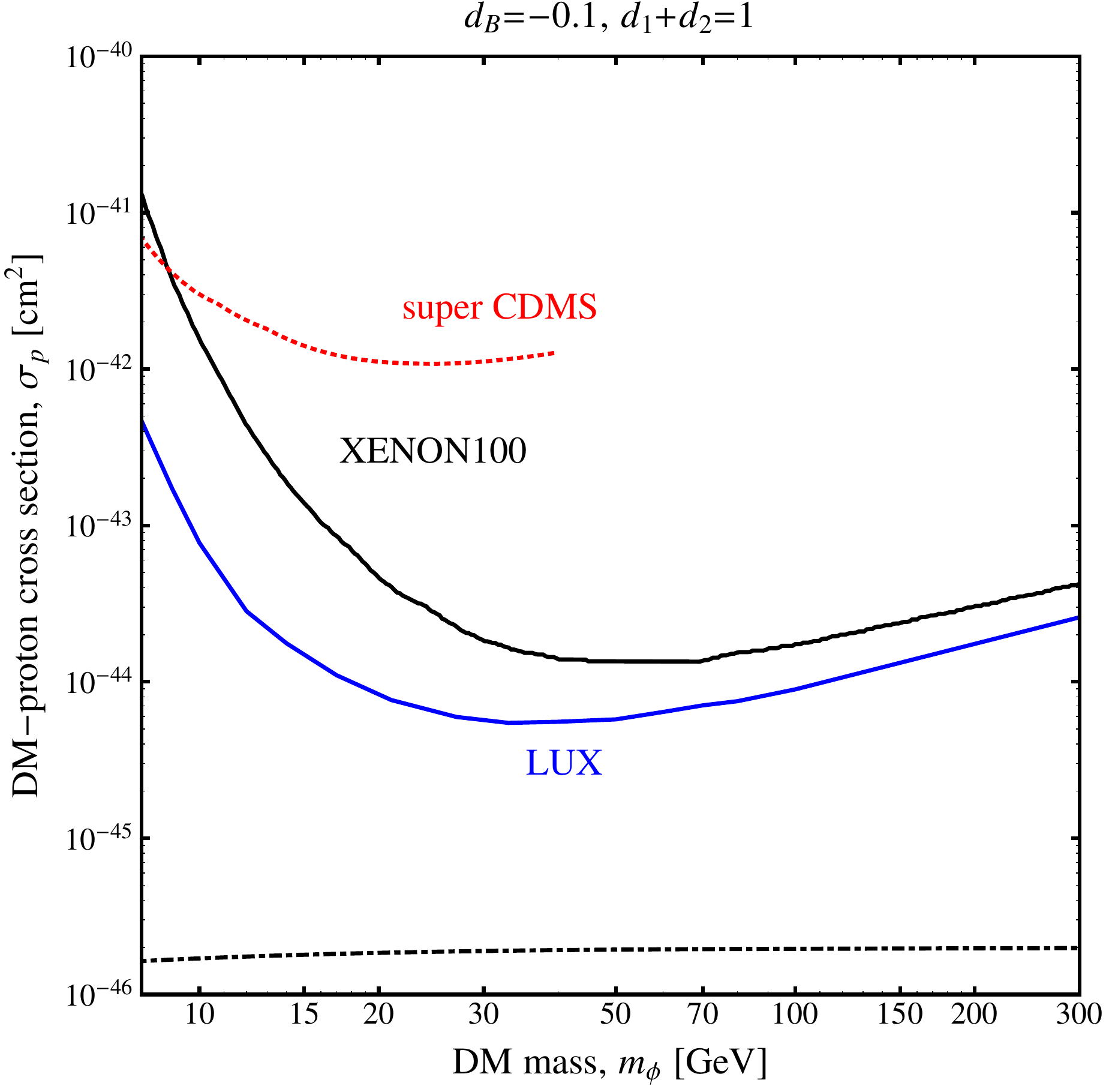}
\caption{\em\label{fig:experiments}Exclusion contours in the $(m_\phi, \sigma_p)$ plane for $d_B= -1$ (top panel) and $d_B=  - 0.1$ (bottom panel).  The red-dotted contour is the exclusion plot by Super CDMS \cite{Agnese:2014aze}; and the black and blue lines are respectively the exclusion plots from Xenon100 \cite{Aprile:2011hi} and LUX \cite{Akerib:2013tjd} experiments. The composite Goldstone DM cross sections are the black-dot-dashed curves for $d_1+d_2 = 1$ in both figures with $d_B=-1$ for the top figure and $d_B=-0.1$ for the bottom one. }
\end{figure}

In the top and bottom panels of Fig.~\ref{fig:experiments} we plot the exclusion limits from Super CDMS, Xenon100 and LUX in the $(m_{\phi},\sigma_p)$ plane for $d_B = -1$ (top panel) and $d_B=-0.1$ (bottom panel). In both cases we used the value $d_1 + d_2 =1$.  
From the figure we observe that LUX and XENON100 start putting interesting constraints on the composite GB DM parameters for masses between 15 and 300 GeVs. In particular we cannot have too large values for the isospin breaking parameter $d_B$. 

\section{Conclusions}\label{sec:conc}

We find that the theoretical composite GB DM cross sections (the black dot-dashed curves of Fig.~\ref{fig:experiments})  is constrained by the most stringent experiments for sufficiently large weak isospin breaking and a composite GB DM mass between 15 and 300 GeVs. The maximal size of the cross section with ordinary matter, at low energies, is set by having explicitly shown, via lattice simulations, that in this theory the relevant form factors are saturated by a single vector meson exchange whose mass is in the $2.5$ TeV energy range. 

If the isospin breaking parameter is small one can envision models with larger cross sections. These would require smaller values of the vector masses which can be obtained, for example,  by rendering the theory near conformal by either adding new matter gauged under the composite dynamics and singlet with respect to SM interactions \cite{Dietrich:2006cm,Ryttov:2008xe}, and/or changing the matter representation or the composite gauge group \cite{Frandsen:2009mi,Hietanen:2012sz}.  Lattice investigations of non-GB composite DM were performed in \cite{Appelquist:2013ms,Appelquist:2014jch,Detmold:2014qqa,Detmold:2014kba}.  

\acknowledgments
We thank Ian Shoemaker for discussions and for adding the latest constraints for DM in Fig.~\ref{fig:experiments}. This work was supported in part by the Natural Sciences and Engineering
Research Council (NSERC) of Canada, the Danish National Research Foundation
DNRF:90 grant, and by a Lundbeck Foundation Fellowship grant.
Computing facilities were provided by the Danish Centre for Scientific Computing and
Canada's Shared Hierarchical Academic Research Computing Network
(SHARCNET: http://www.sharcnet.ca).
\begin{table}
\begin{tabular}{cccc}
\hline
$\beta$ & $m_0$ & $Q^2$ & $F_{\Pi}$ \\
\hline\hline
2.0 & -0.947 & 0 & 1(0)\\
2.0 & -0.947 & 0.03638 & 0.80(3)\\
2.0 & -0.947 & 0.0695 & 0.71(3)\\
2.0 & -0.947 & 0.1002 & 0.62(4)\\
2.0 & -0.947 & 0.1279 & 0.57(6)\\
2.0 & -0.947 & 0.1554 & 0.52(4)\\
2.0 & -0.947 & 0.1815 & 0.46(4)\\
2.0 & -0.947 & 0.2301 & 0.43(5)\\
2.0 & -0.947 & 0.2537 & 0.39(5)\\
2.2 & -0.72 & 0 & 1(0)\\
2.2 & -0.72 & 0.03698 & 0.839(14)\\
2.2 & -0.72 & 0.07154 & 0.73(3)\\
2.2 & -0.72 & 0.1042 & 0.64(3)\\
2.2 & -0.72 & 0.1341 & 0.52(5)\\
2.2 & -0.72 & 0.1639 & 0.49(5)\\
2.2 & -0.72 & 0.1927 & 0.45(4)\\
2.2 & -0.72 & 0.2466 & 0.39(5)\\
2.2 & -0.72 & 0.2729 & 0.37(5)\\
\hline
\end{tabular}
\hspace{40pt}
\begin{tabular}{cccc}
\hline
$\beta$ & $m_0$ & $Q^2$ & $F_{\Pi}$ \\
\hline\hline
2.2 & -0.72 & 0.2689 & 0.34(6)\\
2.2 & -0.72 & 0.2948 & 0.30(6)\\
2.2 & -0.72 & 0.3201 & 0.32(6)\\
2.2 & -0.72 & 0.3476 & 0.32(5)\\
2.2 & -0.72 & 0.3683 & 0.23(6)\\
2.2 & -0.72 & 0.3922 & 0.24(5)\\
2.2 & -0.75 & 0 & 1(0)\\
2.2 & -0.75 & 0.0346 & 0.79(3)\\
2.2 & -0.75 & 0.0639 & 0.64(4)\\
2.2 & -0.75 & 0.09 & 0.58(5)\\
2.2 & -0.75 & 0.113 & 0.50(6)\\
2.2 & -0.75 & 0.1352 & 0.44(6)\\
2.2 & -0.75 & 0.1561 & 0.45(5)\\
2.2 & -0.75 & 0.1942 & 0.30(6)\\
2.2 & -0.75 & 0.2097 & 0.34(8)\\
2.2 & -0.75 & 0.2275 & 0.29(6)\\
2.2 & -0.75 & 0.2447 & 0.32(5)\\
\hline
&&& \\
\end{tabular}
\caption{The values for $F_{\Pi}$.\label{table:fpi}}
\end{table}

\appendix

\section{Lattice measurements}

In this appendix we list the numbers of the lattice measurements of
$F_{\Pi}$ in Table~\ref{table:fpi}.


\begin{thebibliography}{30}
\bibitem{Nussinov:1985xr} 
  S.~Nussinov,
  Phys.\ Lett.\ B {\bf 165}, 55 (1985).

\bibitem{Dietrich:2006cm} 
  D.~D.~Dietrich and F.~Sannino,
  Phys.\ Rev.\ D {\bf 75}, 085018 (2007)
  [hep-ph/0611341].

\bibitem{Nardi:2008ix}
  E.~Nardi, F.~Sannino and A.~Strumia,
  JCAP {\bf 0901} (2009) 043
  [arXiv:0811.4153 [hep-ph]].

\bibitem{Gudnason:2006ug} 
  S.~B.~Gudnason, C.~Kouvaris and F.~Sannino,
  Phys.\ Rev.\ D {\bf 73}, 115003 (2006)
  [hep-ph/0603014].

\bibitem{Foadi:2008qv} 
  R.~Foadi, M.~T.~Frandsen and F.~Sannino,
  Phys.\ Rev.\ D {\bf 80}, 037702 (2009)
  [arXiv:0812.3406 [hep-ph]].

\bibitem{Khlopov:2008ty} 
  M.~Y.~.Khlopov and C.~Kouvaris,
  Phys.\ Rev.\ D {\bf 78}, 065040 (2008)
  [arXiv:0806.1191 [astro-ph]].



\bibitem{Sannino:2009za} 
  F.~Sannino,
  Acta Phys.\ Polon.\ B {\bf 40}, 3533 (2009)
  [arXiv:0911.0931 [hep-ph]].


\bibitem{Ryttov:2008xe} 
  T.~A.~Ryttov and F.~Sannino,
  Phys.\ Rev.\ D {\bf 78}, 115010 (2008)
  [arXiv:0809.0713 [hep-ph]].

\bibitem{Kaplan:2009ag} 
  D.~E.~Kaplan, M.~A.~Luty and K.~M.~Zurek,
  Phys.\ Rev.\ D {\bf 79}, 115016 (2009)
  [arXiv:0901.4117 [hep-ph]].

\bibitem{Frandsen:2009mi} 
  M.~T.~Frandsen and F.~Sannino,
  Phys.\ Rev.\ D {\bf 81}, 097704 (2010)
  [arXiv:0911.1570 [hep-ph]].

\bibitem{Belyaev:2010kp} 
  A.~Belyaev, M.~T.~Frandsen, S.~Sarkar and F.~Sannino,
  Phys.\ Rev.\ D {\bf 83}, 015007 (2011)
  [arXiv:1007.4839 [hep-ph]].

\bibitem{Kaplan:1983fs} 
  D.~B.~Kaplan and H.~Georgi,
  Phys.\ Lett.\ B {\bf 136}, 183 (1984).

\bibitem{Cacciapaglia:2014uja} 
  G.~Cacciapaglia and F.~Sannino,
  JHEP {\bf 1404}, 111 (2014)
  [arXiv:1402.0233 [hep-ph]].

\bibitem{Lewis:2011zb} 
  R.~Lewis, C.~Pica and F.~Sannino,
  Phys.\ Rev.\ D {\bf 85}, 014504 (2012)
  [arXiv:1109.3513 [hep-ph]].

\bibitem{Hietanen:2014xca} 
  A.~Hietanen, R.~Lewis, C.~Pica and F.~Sannino,
  JHEP {\bf 1407}, 116 (2014)
  [arXiv:1404.2794 [hep-lat]].

\bibitem{Foadi:2012bb} 
  R.~Foadi, M.~T.~Frandsen and F.~Sannino,
  Phys.\ Rev.\ D {\bf 87}, 095001 (2013)
  [arXiv:1211.1083 [hep-ph]].

\bibitem{Appelquist:2013sia} 
  T.~Appelquist, R.~Brower, S.~Catterall, G.~Fleming, J.~Giedt, A.~Hasenfratz, J.~Kuti and E.~Neil {\it et al.},
  arXiv:1309.1206 [hep-lat].

\bibitem{Appelquist:2013pqa} 
  T.~Appelquist, R.~C.~Brower, M.~I.~Buchoff, M.~Cheng, G.~T.~Fleming, J.~Kiskis, M.~F.~Lin and E.~T.~Neil {\it et al.},
  Phys.\ Rev.\ Lett.\  {\bf 112}, 111601 (2014)
  [arXiv:1311.4889 [hep-ph]].

\bibitem{Nagai:2009ip} 
  K.~i.~Nagai, G.~Carrillo-Ruiz, G.~Koleva and R.~Lewis,
  Phys.\ Rev.\ D {\bf 80}, 074508 (2009)
  [arXiv:0908.0166 [hep-lat]].

\bibitem{Hands:2006ve} 
  S.~Hands, S.~Kim and J.~I.~Skullerud,
  Eur.\ Phys.\ J.\ C {\bf 48}, 193 (2006)
  [hep-lat/0604004].

\bibitem{Hands:2007uc} 
  S.~Hands, P.~Sitch and J.~I.~Skullerud,
  Phys.\ Lett.\ B {\bf 662}, 405 (2008)
  [arXiv:0710.1966 [hep-lat]].

\bibitem{Hands:2010gd} 
  S.~Hands, S.~Kim and J.~I.~Skullerud,
  Phys.\ Rev.\ D {\bf 81}, 091502 (2010)
  [arXiv:1001.1682 [hep-lat]].

\bibitem{Hands:2011hd} 
  S.~Hands and P.~Kenny,
  Phys.\ Lett.\ B {\bf 701}, 373 (2011)
  [arXiv:1104.0522 [hep-lat]].

\bibitem{Cotter:2012mb} 
  S.~Cotter, P.~Giudice, S.~Hands and J.~I.~Skullerud,
  Phys.\ Rev.\ D {\bf 87}, no. 3, 034507 (2013)
  [arXiv:1210.4496 [hep-lat]].

\bibitem{Matsufuru:2014dva} 
  H.~Matsufuru, Y.~Kikukawa, K.~i.~Nagai and N.~Yamada,
  PoS LATTICE {\bf 2013}, 123 (2014)
  [arXiv:1401.6655 [hep-lat]].

\bibitem{Detmold:2014qqa} 
  W.~Detmold, M.~McCullough and A.~Pochinsky,
  arXiv:1406.2276 [hep-ph].

\bibitem{Detmold:2014kba} 
  W.~Detmold, M.~McCullough and A.~Pochinsky,
  arXiv:1406.4116 [hep-lat].

\bibitem{Appelquist:2013ms} 
  T.~Appelquist {\it et al.}  [Lattice Strong Dynamics (LSD) Collaboration],
  Phys.\ Rev.\ D {\bf 88}, no. 1, 014502 (2013)
  [arXiv:1301.1693 [hep-ph]].

\bibitem{Appelquist:2014jch} 
  T.~Appelquist {\it et al.}  [Lattice Strong Dynamics (LSD) Collaboration],
  arXiv:1402.6656 [hep-lat].

\bibitem{DelNobile:2011je} 
  E.~Del Nobile, C.~Kouvaris and F.~Sannino,
  Phys.\ Rev.\ D {\bf 84}, 027301 (2011)
  [arXiv:1105.5431 [hep-ph]].

\bibitem{DelNobile:2011yb} 
  E.~Del Nobile, C.~Kouvaris, F.~Sannino and J.~Virkajarvi,
  Mod.\ Phys.\ Lett.\ A {\bf 27}, 1250108 (2012)
  [arXiv:1111.1902 [hep-ph]].

\bibitem{Bernabei:2008yi} 
  R.~Bernabei {\it et al.}  [DAMA Collaboration],
  Eur.\ Phys.\ J.\ C {\bf 56}, 333 (2008)
  [arXiv:0804.2741 [astro-ph]].

\bibitem{Aprile:2011hi} 
  E.~Aprile {\it et al.}  [XENON100 Collaboration],
  Phys.\ Rev.\ Lett.\  {\bf 107}, 131302 (2011)
  [arXiv:1104.2549 [astro-ph.CO]].

\bibitem{Aprile:2012nq} 
  E.~Aprile {\it et al.}  [XENON100 Collaboration],
  Phys.\ Rev.\ Lett.\  {\bf 109}, 181301 (2012)
  [arXiv:1207.5988 [astro-ph.CO]].

\bibitem{Ahmed:2009zw} 
  Z.~Ahmed {\it et al.}  [CDMS-II Collaboration],
  Science {\bf 327}, 1619 (2010)
  [arXiv:0912.3592 [astro-ph.CO]].
%
%
%
%
%
%
%


\bibitem{Akerib:2013tjd} 
  D.~S.~Akerib {\it et al.}  [LUX Collaboration],
  Phys.\ Rev.\ Lett.\  {\bf 112}, 091303 (2014)
  [arXiv:1310.8214 [astro-ph.CO]].
  
\bibitem{Agnese:2014aze} 
  R.~Agnese {\it et al.}  [SuperCDMS Collaboration],
  Phys.\ Rev.\ Lett.\  {\bf 112}, 241302 (2014)
  [arXiv:1402.7137 [hep-ex]].
  
\bibitem{Bonnet:2004fr} 
  F.~D.~R.~Bonnet {\it et al.}  [Lattice Hadron Physics Collaboration],
  Phys.\ Rev.\ D {\bf 72}, 054506 (2005)
  [hep-lat/0411028].

\bibitem{Gasser:1984gg} 
  J.~Gasser and H.~Leutwyler,
  Nucl.\ Phys.\ B {\bf 250}, 465 (1985).


\bibitem{Brommel:2006ww} 
  D.~Brommel {\it et al.}  [QCDSF/UKQCD Collaboration],
  Eur.\ Phys.\ J.\ C {\bf 51}, 335 (2007)
  [hep-lat/0608021].

\bibitem{Frezzotti:2008dr} 
  R.~Frezzotti {\it et al.}  [ETM Collaboration],
  Phys.\ Rev.\ D {\bf 79}, 074506 (2009)
  [arXiv:0812.4042 [hep-lat]].

\bibitem{Boyle:2008yd} 
  P.~A.~Boyle, J.~M.~Flynn, A.~Juttner, C.~Kelly, H.~P.~de Lima, C.~M.~Maynard, C.~T.~Sachrajda and J.~M.~Zanotti,
  JHEP {\bf 0807}, 112 (2008)
  [arXiv:0804.3971 [hep-lat]].

\bibitem{Aoki:2009qn} 
  S.~Aoki {\it et al.}  [JLQCD and TWQCD Collaborations],
  Phys.\ Rev.\ D {\bf 80}, 034508 (2009)
  [arXiv:0905.2465 [hep-lat]].

\bibitem{Nguyen:2011ek} 
  O.~H.~Nguyen, K.~-I.~Ishikawa, A.~Ukawa and N.~Ukita,
  JHEP {\bf 1104}, 122 (2011)
  [arXiv:1102.3652 [hep-lat]].

\bibitem{Brandt:2011jk} 
  B.~B.~Brandt, A.~Juttner and H.~Wittig,
  arXiv:1109.0196 [hep-lat].

\bibitem{Brandt:2013dua} 
  B.~B.~Brandt, A.~Jttner and H.~Wittig,
  JHEP {\bf 1311}, 034 (2013)
  [arXiv:1306.2916 [hep-lat]].

\bibitem{Wilcox:1985iy} 
  W.~Wilcox and R.~M.~Woloshyn,
  Phys.\ Rev.\ Lett.\  {\bf 54}, 2653 (1985).

\bibitem{Woloshyn:1985in} 
  R.~M.~Woloshyn and A.~M.~Kobos,
  Phys.\ Rev.\ D {\bf 33}, 222 (1986).

\bibitem{Woloshyn:1985vd} 
  R.~M.~Woloshyn,
  Phys.\ Rev.\ D {\bf 34}, 605 (1986).

\bibitem{'tHooft:1973jz} 
  G.~'t Hooft,
  Nucl.\ Phys.\ B {\bf 72}, 461 (1974).

\bibitem{Witten:1979kh} 
  E.~Witten,
  Nucl.\ Phys.\ B {\bf 160}, 57 (1979).

\bibitem{Masjuan:2012sk} 
  P.~Masjuan, E.~Ruiz Arriola and W.~Broniowski,
  Phys.\ Rev.\ D {\bf 87}, 014005 (2013)
  [arXiv:1210.0760 [hep-ph]].

\bibitem{Beringer:1900zz} 
  J.~Beringer {\it et al.}  [Particle Data Group Collaboration],
  Phys.\ Rev.\ D {\bf 86}, 010001 (2012).

\bibitem{DelDebbio:2008zf} 
  L.~Del Debbio, A.~Patella and C.~Pica,
  Phys.\ Rev.\ D {\bf 81}, 094503 (2010)
  [arXiv:0805.2058 [hep-lat]].

\bibitem{Chang:2010yk} 
  S.~Chang, J.~Liu, A.~Pierce, N.~Weiner and I.~Yavin,
  JCAP {\bf 1008}, 018 (2010)
  [arXiv:1004.0697 [hep-ph]].

\bibitem{Feng:2011vu} 
  J.~L.~Feng, J.~Kumar, D.~Marfatia and D.~Sanford,
  Phys.\ Lett.\ B {\bf 703}, 124 (2011)
  [arXiv:1102.4331 [hep-ph]].

\bibitem{Savage:2010tg} 
  C.~Savage, G.~Gelmini, P.~Gondolo and K.~Freese,
  Phys.\ Rev.\ D {\bf 83}, 055002 (2011)
  [arXiv:1006.0972 [astro-ph.CO]].

\bibitem{Bozorgnia:2010xy} 
  N.~Bozorgnia, G.~B.~Gelmini and P.~Gondolo,
  JCAP {\bf 1011}, 019 (2010)
  [arXiv:1006.3110 [astro-ph.CO]].

\bibitem{Ahmed:2010wy} 
  Z.~Ahmed {\it et al.}  [CDMS-II Collaboration],
  Phys.\ Rev.\ Lett.\  {\bf 106}, 131302 (2011)
  [arXiv:1011.2482 [astro-ph.CO]].

\bibitem{Angle:2011th} 
  J.~Angle {\it et al.}  [XENON10 Collaboration],
  Phys.\ Rev.\ Lett.\  {\bf 107}, 051301 (2011)
  [Erratum-ibid.\  {\bf 110}, 249901 (2013)]
  [arXiv:1104.3088 [astro-ph.CO]].


\bibitem{Hietanen:2012sz} 
  A.~Hietanen, C.~Pica, F.~Sannino and U.~I.~Sondergaard,
  Phys.\ Rev.\ D {\bf 87} 034508 (2013)
  [arXiv:1211.5021 [hep-lat]].

\end{thebibliography}
\end{document}